\documentclass[vecphys]{svmult}

\usepackage{makeidx}         
\usepackage{graphicx}        
\usepackage{multicol}        
\usepackage{cite}            
\usepackage[bottom]{footmisc}

\makeindex             

\begin{document}

\title{Electronic properties of monolayer and bilayer graphene}
\author{Edward McCann}
\institute{Department of Physics, Lancaster University, Lancaster, LA1
4YB, United Kingdom
\texttt{ed.mccann@lancaster.ac.uk}}

\setcounter{page}{1}

\maketitle

\begin{abstract}
The tight-binding model of electrons in graphene is reviewed.
We derive low-energy Hamiltonians supporting massless
Dirac-like chiral fermions and massive chiral fermions
in monolayer and bilayer graphene, respectively,
and we describe how their chirality is manifest in the
sequencing of plateaus observed in the integer quantum Hall effect.
The opening of a tuneable band gap in bilayer graphene in response to a
transverse electric field is described, and we explain how Hartree theory
may be used to develop a simple analytical model of screening.
\end{abstract}

\section{Introduction}
\label{s:mcc:intro}

More than sixty years ago, Wallace \cite{wallace46} modeled
the electronic band structure of graphene.
Research into graphene was stimulated by interest
in the properties of bulk graphite because, from a theoretical point of view,
two-dimensional graphene serves as a building block for the three-dimensional
material. Following further work, the tight-binding model of electrons in
graphite, that takes into account coupling between layers, became known as the Slonczewski-Weiss-McClure model \cite{sw58,mcclure57,dressel02}.
As well as serving as the basis for models of carbon-based
materials including graphite, buckyballs, and carbon nanotubes \cite{d+m84,gon92,a+a93,k+m97,ando98,mceuen99,saito},
the honeycomb lattice of graphene has been used theoretically
to study Dirac fermions in a condensed matter system
\cite{semenoff84,haldane88}.
Since the experimental isolation of individual graphene flakes \cite{novo04},
and the observation of the integer quantum Hall effect
in monolayers \cite{novo05,zhang05} and bilayers \cite{novo06},
there has been an explosion of interest in the behavior of
chiral electrons in graphene.

This Chapter begins in Sect.~\ref{s:mcc:crystal} with a description of
the crystal structure of monolayer graphene.
Section~\ref{s:mcc:tbmgeneral} briefly reviews the tight-binding model
of electrons in condensed matter materials \cite{ash+mer,saito},
and Sect.~\ref{s:mcc:tbmmonolayer} describes its application
to monolayer graphene \cite{saito,cnreview,bena09}.
Then, in Section~\ref{s:mcc:chiral}, we explain how a
Dirac-like Hamiltonian describing massless chiral fermions
emerges from the tight-binding model at low energy.
The tight-binding model is applied to bilayer graphene in
Sect.~\ref{s:mcc:tbmbilayer}, and Sect.~\ref{s:mcc:massive} describes
how low-energy electrons in bilayers behave as massive
chiral quasiparticles \cite{novo06,mcc06a}.
In Sect.~\ref{s:mcc:qhe}, we describe how the
chiral Hamiltonians of monolayer and bilayer graphene
corresponding to Berry's phase $\pi$ and $2\pi$, respectively,
have associated four- and eight-fold degenerate zero-energy Landau levels,
leading to an unusual sequence of plateaus in the integer quantum
Hall effect \cite{novo05,zhang05,novo06}.

Section~\ref{s:mcc:twg} discusses an additional contribution to the
low-energy Hamiltonians of monolayer and bilayer graphene, known
as trigonal warping \cite{dressel02,dre74,nak76,ino62,gup72,ando98,mcc06a},
that produces a Liftshitz transition in the band
structure of bilayer graphene at low energy.
Finally, Sect.~\ref{s:mcc:tune} describes how an external transverse electric
field applied to bilayer graphene, due to doping or gates,
may open a band gap that can be tuned between zero up to the value of the interlayer
coupling, around three to four hundred meV \cite{mcc06a,ohta06,oostinga}.
Hartree theory and the tight-binding model are used to develop
a simple model of screening by electrons in bilayer graphene in order
to calculate the density dependence of the band gap \cite{mcc06b}.

\section{The crystal structure of monolayer graphene}
\label{s:mcc:crystal}

\subsection{The real space structure}
\label{ss:mcc:real}

\begin{figure}[t]
\centering
\includegraphics*[width=.6\textwidth]{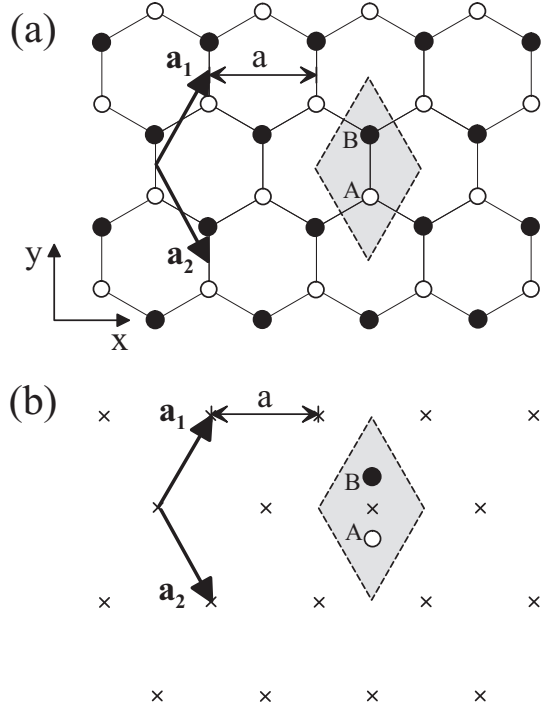}
\caption[]{(a) The honeycomb crystal structure of monolayer graphene where white
(black) circles indicate carbon atoms on $A$ ($B$) sites
and straight lines indicate $\sigma$ bonds between them.
Vectors $\mathbf{a}_1$ and $\mathbf{a}_2$ are primitive lattice
vectors of length equal to the lattice constant $a$.
The shaded rhombus is a unit cell containing
two atoms, one $A$ and one $B$.
(b) Crosses indicate lattice points of the hexagonal Bravais lattice.
The honeycomb structure in (a) consists of the hexagonal Bravais lattice [shown in (b)]
with a basis of two atoms, one $A$ and one $B$, at each lattice point.}
\label{mccfig:1}       
\end{figure}

Monolayer graphene consists of carbon atoms arranged with a
two-dimensional honeycomb
crystal structure as shown in Fig.~\ref{mccfig:1}(a).
The honeycomb structure \cite{ash+mer,saito} consists of the hexagonal
Bravais lattice, Fig.~\ref{mccfig:1}(b),
with a basis of two atoms, labeled $A$ and $B$, at each lattice point.

Throughout this Chapter, we use a Cartesian coordinate system with $x$ and $y$ axes
in the plane of the graphene crystal, and a $z$ axis perpendicular
to the graphene plane. Two-dimensional vectors in the same plane
as the graphene are expressed solely in terms of their $x$ and $y$
coordinates, so that, for example, the primitive lattice vectors
of the hexagonal Bravais lattice, Fig.~\ref{mccfig:1}(b),
are $\mathbf{a}_1$ and $\mathbf{a}_2$ where
\begin{eqnarray}
\mathbf{a}_1 = \left( \frac{a}{2} , \frac{\sqrt{3}a}{2} \right) \, , \qquad
\mathbf{a}_2 = \left( \frac{a}{2} , - \frac{\sqrt{3}a}{2} \right) \, ,
\end{eqnarray}
and $a = | \mathbf{a}_1 | = | \mathbf{a}_2 |$ is the lattice constant.
In graphene, $a = 2.46\,$\AA \cite{saito}.
The lattice constant is the distance between unit cells,
whereas the distance between carbon atoms is the carbon-carbon bond length
$a_{CC} = a / \sqrt{3} = 1.42\,$\AA.
Note that the honeycomb structure is not a Bravais lattice because
atomic positions $A$ and $B$ are not equivalent:
it is not possible to connect them with a lattice vector
$\mathbf{R} = n_1 \mathbf{a}_1 + n_2 \mathbf{a}_2$
where $n_1$ and $n_2$ are integers.
Taken alone, the $A$ atomic positions (or, the $B$ atomic positions)
make up an hexagonal Bravais lattice and, in the following, we will
often refer to them as the `$A$ sublattice' (or, the `$B$ sublattice').

\subsection{The reciprocal lattice of graphene}
\label{ss:mcc:reciplattice}

\begin{figure}[t]
\centering
\includegraphics*[width=.6\textwidth]{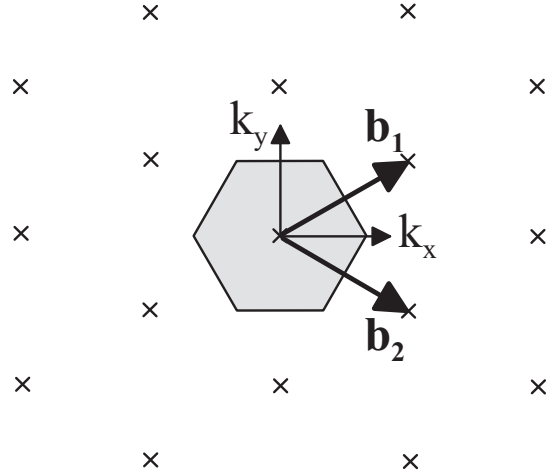}
\caption[]{The reciprocal lattice of monolayer graphene
where crosses indicate reciprocal lattice points,
and vectors $\mathbf{b}_1$ and $\mathbf{b}_2$ are primitive lattice
vectors. The shaded hexagon indicates the first Brillouin zone.}
\label{mccfig:2}
\end{figure}

Primitive reciprocal lattice vectors
$\mathbf{b}_1$ and $\mathbf{b}_2$ satisfying
$\mathbf{a}_1 \mathbf{b}_1 = \mathbf{a}_2 \mathbf{b}_2 = 2 \pi$
and $\mathbf{a}_1 \mathbf{b}_2 = \mathbf{a}_2 \mathbf{b}_1 = 0$
are given by
\begin{eqnarray}
\mathbf{b}_1 = \left( \frac{2\pi}{a} , \frac{2\pi}{\sqrt{3}a} \right) \, , \qquad
\mathbf{b}_2 = \left( \frac{2\pi}{a} , - \frac{2\pi}{\sqrt{3}a} \right) \, . \label{mcc:b1b2}
\end{eqnarray}
The resulting reciprocal lattice is shown in Fig.~\ref{mccfig:2},
which is an hexagonal Bravais lattice. The first Brillouin zone
is hexagonal, as indicated by the shaded region in Fig.~\ref{mccfig:2}.

\subsection{The atomic orbitals of graphene}
\label{ss:mcc:orbitals}

Each carbon atom has six electrons, of which two
are core electrons and four are valence electrons. The latter
occupy $2s$, $2p_x$, $2p_y$, and $2p_z$ orbitals.
In graphene, the orbitals  are $sp^2$ hybridized, meaning that
two of the $2p$ orbitals, the $2p_x$ and $2p_y$ that lie in the graphene plane,
mix with the $2s$ orbital to form three $sp^2$ hybrid orbitals
per atom, each lying in the graphene plane and oriented $120^{\circ}$ to each other \cite{saito}.
They form $\sigma$ bonds with other atoms, shown as straight
lines in the honeycomb crystal structure, Fig.~\ref{mccfig:1}(a).
The remaining $2p_z$ orbital for each atom lies perpendicular to
the plane, and, when combined with the $2p_z$ orbitals on adjacent
atoms in graphene, forms a $\pi$ orbital. Electronic states close
to the Fermi level in graphene are described
well by a model taking into account only the $\pi$ orbital, meaning
that the tight-binding model can include only one electron per atomic site,
in a $2p_z$ orbital.

\section{The tight-binding model}
\label{s:mcc:tbmgeneral}

We begin by presenting a general description of the tight-binding model
for a system with $n$ atomic orbitals $\phi_j$ in the unit cell,
labeled by index $j = 1 \ldots n$. Further details may be found
in the book by Saito, Dresselhaus, and Dresselhaus \cite{saito}.
It is assumed that the system has translational invariance.
Then, the model may be written using $n$ different
Bloch functions $\Phi_j ( \mathbf{k} , \mathbf{r} )$
that depend on the position vector $\mathbf{r}$ and wave vector $\mathbf{k}$.
They are given by
\begin{eqnarray}
\Phi_j ( \mathbf{k} , \mathbf{r} ) = \frac{1}{\sqrt{N}}
\sum_{i=1}^{N} e^{i \mathbf{k}.\mathbf{R}_{j,i}}
\phi_j \left( \mathbf{r} - \mathbf{R}_{j,i} \right) \, , \label{mcc:bloch}
\end{eqnarray}
where the sum is over $N$ different unit cells, labeled by index $i = 1 \ldots N$,
and $\mathbf{R}_{j,i}$ denotes the position of the $j$th orbital in the $i$th
unit cell.

In general, an electronic wave function
$\Psi_j ( \mathbf{k} , \mathbf{r} )$ is given
by a linear superposition of the $n$ different Bloch functions,
\begin{eqnarray}
\Psi_j ( \mathbf{k} , \mathbf{r} ) = \sum_{l=1}^{n}
{c}_{j,l} ( \mathbf{k} ) \, \Phi_l ( \mathbf{k} , \mathbf{r} ) \, , \label{mcc:exp}
\end{eqnarray}
where ${c}_{j,l}$ are coefficients of the expansion.
The energy $E_j (\mathbf{k})$ of the $j$th band is given by
\begin{eqnarray}
E_j (\mathbf{k}) = \frac{\langle \Psi_j | {\cal H} | \Psi_j \rangle}{\langle \Psi_j | \Psi_j \rangle} \, ,
\end{eqnarray}
where ${\cal H}$ is the Hamiltonian.
Substituting the expansion of the wave function
(\ref{mcc:exp}) into the energy gives
\begin{eqnarray}
E_j (\mathbf{k}) &=&
\frac{\sum_{i,l}^{n} {c}_{ji}^{\ast} {c}_{jl} \langle \Phi_i | {\cal H} | \Phi_l \rangle}
{\sum_{i,l}^{n} {c}_{ji}^{\ast} {c}_{jl} \langle \Phi_i | \Phi_l \rangle} \, , \\
&=& \frac{\sum_{i,l}^{n} H_{il} {c}_{ji}^{\ast} {c}_{jl}}
{\sum_{i,l}^{n} S_{il} {c}_{ji}^{\ast} {c}_{jl} } \, , \label{mcc:ej}
\end{eqnarray}
where transfer integral matrix elements $H_{il}$
and overlap integral matrix elements $S_{il}$ are defined by
\begin{eqnarray}
H_{il} = \langle \Phi_i | {\cal H} | \Phi_l \rangle \, , \qquad
S_{il} = \langle \Phi_i | \Phi_l \rangle \, . \label{mcc:ti}
\end{eqnarray}
We minimize the energy $E_j$ with respect to the coefficient
${c}_{jm}^{\ast}$ by calculating the derivative,
\begin{eqnarray}
\frac{\partial E_j}{\partial {c}_{jm}^{\ast}}
= \frac{\sum_{l}^{n} H_{ml} {c}_{jl}}
{\sum_{i,l}^{n} S_{il} {c}_{ji}^{\ast} {c}_{jl} }
-
\frac{\sum_{i,l}^{n} H_{il} {c}_{ji}^{\ast} {c}_{jl}
\sum_{l}^{n} S_{ml} {c}_{jl}}
{\left(\sum_{i,l}^{n} S_{il} {c}_{ji}^{\ast} {c}_{jl}\right)^2} \, .
\end{eqnarray}
The second term contains a factor equal to the energy $E_j$
itself, (\ref{mcc:ej}). Then, setting
$\partial E_j / \partial {c}_{jm}^{\ast} = 0$ and
omitting the common factor $\sum_{i,l}^{n} S_{il} {c}_{ji}^{\ast} {c}_{jl}$ gives
\begin{eqnarray}
\sum_{l=1}^{n} H_{ml} {c}_{jl} = E_j \sum_{l=1}^{n} S_{ml} {c}_{jl} \, . \label{mcc:sum}
\end{eqnarray}
This can be written as a matrix equation. Consider the specific example of
two orbitals per unit cell, $n=2$. Then, we can select the possible values of
$m$ (either $m=1$ or $m=2$) and write out the summation in
(\ref{mcc:sum}) explicitly:
\begin{eqnarray}
m=1 \quad \Rightarrow \quad H_{11} {c}_{j1} + H_{12} {c}_{j2} &=&
E_j \left( S_{11} {c}_{j1} + S_{12} {c}_{j2} \right) \, , \\
m=2 \quad \Rightarrow \quad H_{21} {c}_{j1} + H_{22} {c}_{j2} &=&
E_j \left( S_{21} {c}_{j1} + S_{22} {c}_{j2} \right) \, .
\end{eqnarray}
These two equations may be combined into a matrix equation,
\begin{eqnarray}
\left(
\begin{array}{cc}
H_{11} & H_{12} \\
H_{21} & H_{22} \\
\end{array}
\right)
\left(
  \begin{array}{c}
    {c}_{j1} \\
    {c}_{j2} \\
  \end{array}
\right)
&=& E_j
\left(
\begin{array}{cc}
S_{11} & S_{12} \\
S_{21} & S_{22} \\
\end{array}
\right)
\left(
  \begin{array}{c}
    {c}_{j1} \\
    {c}_{j2} \\
  \end{array}
\right) \, .
\end{eqnarray}
For general values of $n$, defining $H$ as
the transfer integral matrix, $S$ as the
overlap integral matrix and $\psi_j$ as a
column vector,
\begin{eqnarray}
\!\!\!\!\!\!\!\!\!\!\!\!H = \left(
      \begin{array}{cccc}
        H_{11} & H_{12} & \cdots & H_{1n} \\
        H_{21} & H_{22} & \cdots & H_{2n} \\
        \vdots & \vdots & \ddots & \vdots \\
        H_{n1} & H_{n2} & \cdots & H_{nn} \\
      \end{array}
    \right) \!\! , \,
S = \left(
      \begin{array}{cccc}
        S_{11} & S_{12} & \cdots & S_{1n} \\
        S_{21} & S_{22} & \cdots & S_{2n} \\
        \vdots & \vdots & \ddots & \vdots \\
        S_{n1} & S_{n2} & \cdots & S_{nn} \\
      \end{array}
    \right) \!\! , \,
\psi_j = \left(
           \begin{array}{c}
             {c}_{j1} \\
             {c}_{j2} \\
             \vdots \\
             {c}_{jn} \\
           \end{array}
         \right) \!\! , \label{mcc:HSpsi}
\end{eqnarray}
allows the relation (\ref{mcc:sum}) to be expressed as
\begin{eqnarray}
H \psi_j &=& E_j S \psi_j \, .   \label{mcc:HES}
\end{eqnarray}
The energies $E_j$ may be determined by
solving the secular equation
\begin{eqnarray}
\det \left( H -  E_j S \right) = 0 \, , \label{mcc:sec}
\end{eqnarray}
once the transfer integral matrix $H$ and the
overlap integral matrix $S$ are known.
Here, `$\det$' stands for the determinant of the matrix.
In the following, we will omit the subscript $j = 1 \ldots n$
in (\ref{mcc:HES}),(\ref{mcc:sec}), bearing in mind that the number
of solutions is equal to the number of different atomic orbitals
per unit cell.

\section{The tight-binding model of monolayer graphene}
\label{s:mcc:tbmmonolayer}

We apply the tight-binding model described in Sect.~\ref{s:mcc:tbmgeneral}
to monolayer graphene, taking into account one $2p_z$ orbital per atomic site.
As there are two atoms in the unit cell of graphene, labeled $A$ and $B$
in Fig.~\ref{mccfig:1}, the model includes two Bloch functions, $n=2$.
For simplicity, we replace index $j=1$ with $j=A$, and $j=2$ with $j=B$.
Now we proceed to determine the transfer integral matrix $H$ and the
overlap integral matrix $S$.

\subsection{Diagonal matrix elements}
\label{ss:mcc:diag}

Substituting the expression for the Bloch function (\ref{mcc:bloch})
into the definition of the transfer integral (\ref{mcc:ti})
allows us to write the diagonal matrix element corresponding to the $A$
sublattice as
\begin{eqnarray}
\!\!\!\!\!\! H_{AA} = \frac{1}{N} \sum_{i=1}^{N} \sum_{j=1}^{N}
e^{i \mathbf{k}. \left( \mathbf{R}_{A,j} - \mathbf{R}_{A,i} \right)}
\langle \phi_A \left( \mathbf{r} - \mathbf{R}_{A,i} \right) | {\cal H} | \phi_A \left( \mathbf{r} - \mathbf{R}_{A,j} \right) \rangle  , \label{mcc:diagsum}
\end{eqnarray}
where $\mathbf{k} = ( k_x , k_y )$ is the wave vector in the graphene plane.
Equation~(\ref{mcc:diagsum}) includes a double summation over all the $A$ sites of the lattice.
If we assume that the dominant contribution arises from the same site $j = i$
within every unit cell, then:
\begin{eqnarray}
H_{AA} \approx \frac{1}{N} \sum_{i=1}^{N}
\langle \phi_A \left( \mathbf{r} - \mathbf{R}_{A,i} \right) | {\cal H} | \phi_A \left( \mathbf{r} - \mathbf{R}_{A,i} \right) \rangle \, ,
\end{eqnarray}
The matrix element $\langle \phi_A | {\cal H} | \phi_A \rangle$ within the summation
has the same value on every $A$ site, {i.e.} it is independent
of the site index $i$. We set it to be equal to a parameter
\begin{eqnarray}
\epsilon_{2p} =
\langle \phi_A \left( \mathbf{r} - \mathbf{R}_{A,i} \right) | {\cal H} | \phi_A \left( \mathbf{r} - \mathbf{R}_{A,i} \right) \rangle \, ,
\end{eqnarray}
that is equal to the energy of the $2p_z$ orbital. Then, keeping only
the same site contribution,
\begin{eqnarray}
H_{AA} \approx \frac{1}{N} \sum_{i=1}^{N}
\epsilon_{2p} = \epsilon_{2p} \, .
\end{eqnarray}
It is possible to take into account the contribution of other terms in
the double summation (\ref{mcc:diagsum}), such as next-nearest
neighbor contributions \cite{sasaki,peres06}. They generally have a small effect on the
electronic band structure and will not be discussed here.
The $B$ sublattice has the same structure as the $A$ sublattice,
and the carbon atoms on the two sublattices are chemically identical. This means
that the diagonal transfer integral matrix element corresponding to the $B$
sublattice has the same value as that of the $A$ sublattice:
\begin{eqnarray}
H_{BB} = H_{AA} \approx \epsilon_{2p} \, . \label{mcc:Hdiag}
\end{eqnarray}

A calculation of the diagonal elements of the overlap integral matrix proceeds
in a similar way as for those of the transfer integral. In this case,
the overlap between a $2p_z$ orbital on the same atom is equal to unity,
\begin{eqnarray}
\langle \phi_A \left( \mathbf{r} - \mathbf{R}_{A,i} \right) | \phi_A \left( \mathbf{r} - \mathbf{R}_{A,i} \right) \rangle = 1 \, .
\end{eqnarray}
Then, assuming that the same site contribution dominates,
\begin{eqnarray}
S_{AA} &=& \frac{1}{N} \sum_{i=1}^{N} \sum_{j=1}^{N}
e^{i \mathbf{k}. \left( \mathbf{R}_{A,j} - \mathbf{R}_{A,i} \right)}
\langle \phi_A \left( \mathbf{r} - \mathbf{R}_{A,i} \right) | \phi_A \left( \mathbf{r} - \mathbf{R}_{A,j} \right) \rangle \, , \label{mcc:Sdiagsum} \\
&\approx& \frac{1}{N} \sum_{i=1}^{N}
\langle \phi_A \left( \mathbf{r} - \mathbf{R}_{A,i} \right) | \phi_A \left( \mathbf{r} - \mathbf{R}_{A,i} \right) \rangle \, , \\
&=& \frac{1}{N} \sum_{i=1}^{N} 1 \\
&=& 1 \, .
\end{eqnarray}
Again, as the $B$ sublattice has the same structure as the $A$ sublattice,
\begin{eqnarray}
S_{BB} = S_{AA} = 1 \, . \label{mcc:Sdiag}
\end{eqnarray}

\subsection{Off-diagonal matrix elements}
\label{ss:mcc:offdiag}

Substituting the expression for the Bloch function (\ref{mcc:bloch})
into the definition of the transfer integral (\ref{mcc:ti})
allows us to write an off-diagonal matrix element as
\begin{eqnarray}
\!\!\!\!\!\! H_{AB} = \frac{1}{N} \sum_{i=1}^{N} \sum_{j=1}^{N}
e^{i \mathbf{k}. \left( \mathbf{R}_{B,j} - \mathbf{R}_{A,i} \right)}
\langle \phi_A \left( \mathbf{r} - \mathbf{R}_{A,i} \right) | {\cal H} | \phi_B \left( \mathbf{r} - \mathbf{R}_{B,j} \right) \rangle . \label{mcc:offdiagsum}
\end{eqnarray}
It describes processes of hopping between the $A$ and $B$ sublattices,
and contains a summation over all the $A$ sites ($i=1 \ldots N$)
at positions $\mathbf{R}_{A,i}$ and all the $B$ sites ($j=1 \ldots N$)
at $\mathbf{R}_{B,j}$.

\begin{figure}[t]
\centering
\includegraphics*[width=.6\textwidth]{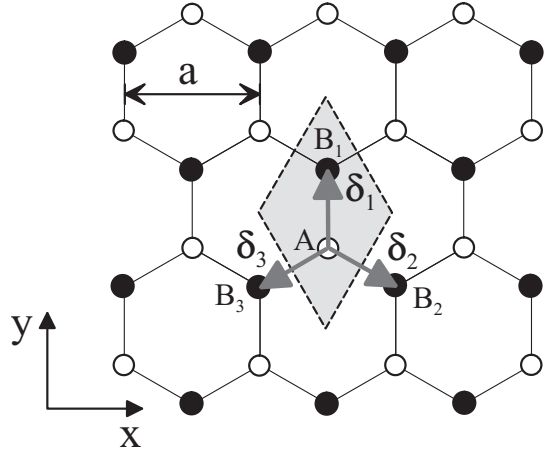}
\caption[]{The honeycomb crystal structure of monolayer graphene.
In the nearest-neighbor approximation, we consider hopping
from an $A$ site (white) to three adjacent $B$ sites (black),
labeled $B_1$, $B_2$, $B_3$, with position vectors
$\mbox{\boldmath$\delta$}_1$, $\mbox{\boldmath$\delta$}_2$,
$\mbox{\boldmath$\delta$}_3$, respectively, relative to the $A$ site.}
\label{mccfig:3}
\end{figure}

In the following, we assume that the dominant
contribution to the off-diagonal matrix element (\ref{mcc:offdiagsum})
arises from hopping between nearest neighbors only.
If we focus on an individual $A$ atom, {i.e.} we consider a
fixed value of the index $i$, we see that it has three neighboring
$B$ atoms, Fig.~\ref{mccfig:3}, that we will label with a
new index $l$ ($l=1 \ldots 3$). Each $A$ atom has three such neighbors,
so it is possible to write the nearest-neighbors contribution to the off-diagonal
matrix element (\ref{mcc:offdiagsum}) as
\begin{eqnarray}
\!\!\! H_{AB} \approx \frac{1}{N} \sum_{i=1}^{N} \sum_{l=1}^{3}
e^{i \mathbf{k}. \left( \mathbf{R}_{B,l} - \mathbf{R}_{A,i} \right)}
\langle \phi_A \left( \mathbf{r} - \mathbf{R}_{A,i} \right) | {\cal H} | \phi_B \left( \mathbf{r} - \mathbf{R}_{B,l} \right) \rangle \, .
\end{eqnarray}
The matrix element between neighboring atoms,
$\langle \phi_A | {\cal H} | \phi_B \rangle$,
has the same value for each neighboring pair,
{i.e.} it is independent of indices $i$ and $l$.
We set it equal to a parameter,
$t = \langle \phi_A \left( \mathbf{r} - \mathbf{R}_{A,i} \right) | {\cal H} |
\phi_B \left( \mathbf{r} - \mathbf{R}_{B,l} \right) \rangle$.
Since $t$ is negative \cite{saito}, it is common practice to express it
in terms of a positive parameter $\gamma_0 = - t$, where
\begin{eqnarray}
\gamma_0 = -
\langle \phi_A \left( \mathbf{r} - \mathbf{R}_{A,i} \right) | {\cal H} | \phi_B \left( \mathbf{r} - \mathbf{R}_{B,l} \right) \rangle \, .
\end{eqnarray}
Then, we write the off-diagonal
transfer integral matrix element as
\begin{eqnarray}
H_{AB} &\approx& - \frac{1}{N} \sum_{i=1}^{N} \sum_{l=1}^{3}
e^{i \mathbf{k}. \left( \mathbf{R}_{B,l} - \mathbf{R}_{A,i} \right)}
\gamma_0 \, , \\
&=& - \frac{\gamma_0}{N} \sum_{i=1}^{N} \sum_{l=1}^{3}
e^{i \mathbf{k}. \mbox{\boldmath$\delta$}_l}
\equiv - \gamma_0 f \left( \mathbf{k} \right) \, , \\
f \left( \mathbf{k} \right) &=& \sum_{l=1}^{3} e^{i \mathbf{k}. \mbox{\boldmath$\delta$}_l} \, ,
\end{eqnarray}
where the position vector of atom $B_l$ relative to the
$A_i$ atom is denoted $\mbox{\boldmath$\delta$}_l = \mathbf{R}_{B,l} - \mathbf{R}_{A,i}$,
and we used the fact that the summation over the three neighboring
$B$ atoms is the same for all $A_i$ atoms.

For the three $B$ atoms shown in Fig.~\ref{mccfig:3}, the three vectors are
\begin{eqnarray}
\mbox{\boldmath$\delta$}_1 = \left( 0 , \frac{a}{\sqrt{3}}\right) \, , \quad
\mbox{\boldmath$\delta$}_2 = \left( \frac{a}{2} , -\frac{a}{2\sqrt{3}}\right) \, , \quad
\mbox{\boldmath$\delta$}_3 = \left( - \frac{a}{2} , -\frac{a}{2\sqrt{3}}\right) \,
\, .
\end{eqnarray}
Note that $|\mbox{\boldmath$\delta$}_1| = |\mbox{\boldmath$\delta$}_2|
= |\mbox{\boldmath$\delta$}_3| = a / \sqrt{3}$ is the carbon-carbon bond length.
Then, the function $f \left( \mathbf{k} \right)$ describing
nearest-neighbor hopping may be evaluated as
\begin{eqnarray}
f \left( \mathbf{k} \right) &=& \sum_{l=1}^{3} e^{i \mathbf{k}. \mbox{\boldmath$\delta$}_l} \, , \label{mcc:fk0} \\
&=& e^{i k_y a /\sqrt{3}} + e^{i k_x a /2} e^{- i k_y a /2\sqrt{3}}
+ e^{-i k_x a /2} e^{- i k_y a /2\sqrt{3}} \, , \\
&=& e^{i k_y a /\sqrt{3}} + 2 e^{- i k_y a /2\sqrt{3}} \cos \left( k_x a / 2 \right) \, . \label{mcc:fk}
\end{eqnarray}
The other off-diagonal matrix element $H_{BA}$ is the complex conjugate of $H_{AB}$:
\begin{eqnarray}
H_{AB} \approx - \gamma_0 f \left( \mathbf{k} \right) \, , \qquad
H_{BA} \approx - \gamma_0 f^{\ast} \left( \mathbf{k} \right) \, . \label{mcc:Hoffdiag}
\end{eqnarray}

A calculation of an off-diagonal element of the overlap integral matrix proceeds
in a similar way as for the transfer integral:
\begin{eqnarray}
\!\!\! S_{AB} &=& \frac{1}{N} \sum_{i=1}^{N} \sum_{j=1}^{N}
e^{i \mathbf{k}. \left( \mathbf{R}_{B,j} - \mathbf{R}_{A,i} \right)}
\langle \phi_A \left( \mathbf{r} - \mathbf{R}_{A,i} \right) | \phi_B \left( \mathbf{r} - \mathbf{R}_{B,j} \right) \rangle , \\
&\approx&
\frac{1}{N} \sum_{i=1}^{N} \sum_{l=1}^{3}
e^{i \mathbf{k}. \left( \mathbf{R}_{B,l} - \mathbf{R}_{A,i} \right)}
\langle \phi_A \left( \mathbf{r} - \mathbf{R}_{A,i} \right) | \phi_B \left( \mathbf{r} - \mathbf{R}_{B,l} \right) \rangle \, , \\
&=& s_0 f \left( \mathbf{k} \right) \, , \label{mcc:Soffdiag}
\end{eqnarray}
where the parameter $s_0 = \langle \phi_A \left( \mathbf{r} - \mathbf{R}_{A,i} \right) | \phi_B \left( \mathbf{r} - \mathbf{R}_{B,l} \right) \rangle$,
and $S_{BA} = S_{AB}^{\ast} = s_0 f^{\ast} \left( \mathbf{k} \right)$.
The presence of non-zero $s_0$ takes into account the possibility
that orbitals on adjacent atomic sites are not strictly orthogonal.

\subsection{The low-energy electronic bands of monolayer graphene}
\label{ss:mcc:bands}

\begin{figure}[t]
\centering
\includegraphics*[width=0.8\textwidth]{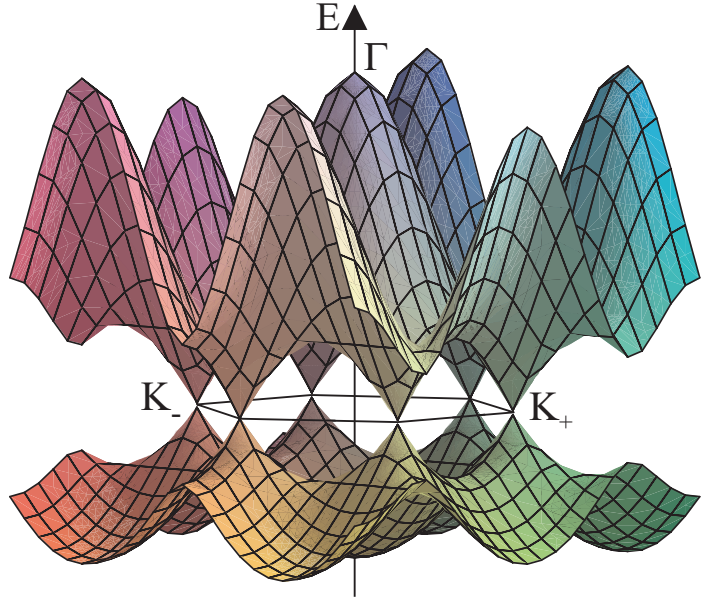}
\caption[]{The low-energy band structure of monolayer
graphene Eq.~(\ref{mcc:monofull}) taking into account nearest-neighbor hopping
with parameter $\gamma_0 = 3.033\,$eV, nearest-neighbor overlap
parameter $s_0 = 0.129$, and orbital energy $\epsilon_{2p} = 0$ \cite{saito}.
The plot shows the bands calculated in the vicinity of the first Brillouin zone,
with conduction and valence bands touching at six corners of the Brillouin zone,
two of them are labeled $K_{+}$ and $K_{-}$. Label $\Gamma$ indicates
the center of the Brillouin zone.}
\label{mccfig:17}
\end{figure}

\begin{figure}[t]
\centering
\includegraphics*[width=.7\textwidth]{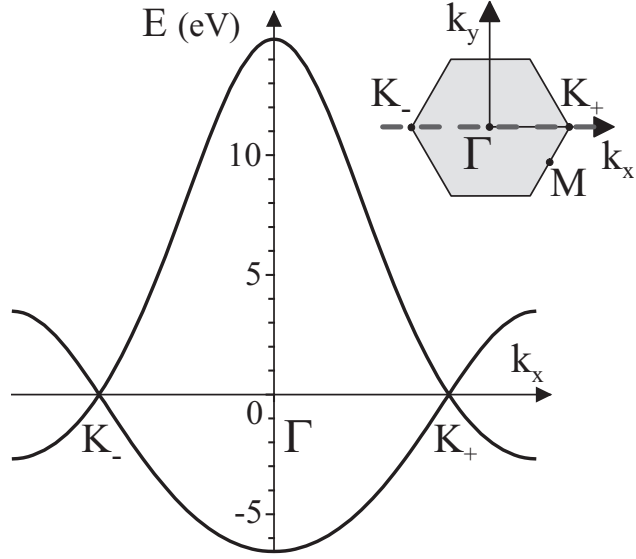}
\caption[]{The low-energy band structure of monolayer
graphene Eq.~(\ref{mcc:monofull}) taking into account nearest-neighbor hopping
with parameter $\gamma_0 = 3.033\,$eV, nearest-neighbor overlap
parameter $s_0 = 0.129$, and orbital energy $\epsilon_{2p} = 0$ \cite{saito}.
The plot shows a cut through the band structure Fig.~\ref{mccfig:17},
plotted along the $k_x$ axis intersecting points $K_{-}$, $\Gamma$, and $K_{+}$ in the Brillouin zone,
shown as the dotted line in the inset.}
\label{mccfig:4}
\end{figure}

Summarizing the results of this section,
the transfer integral matrix elements (\ref{mcc:Hdiag})
and (\ref{mcc:Hoffdiag}), and the overlap integral matrix elements (\ref{mcc:Sdiag})
and (\ref{mcc:Soffdiag}) give
\begin{eqnarray}
H_1 = \left(
      \begin{array}{cc}
        \epsilon_{2p} & - \gamma_0 f \left( \mathbf{k} \right) \\
        - \gamma_0 f^{\ast} \left( \mathbf{k} \right) & \epsilon_{2p} \\
      \end{array}
    \right) \, , \qquad
S_1 = \left(
      \begin{array}{cc}
        1 & s_0 f \left( \mathbf{k} \right) \\
        s_0 f^{\ast} \left( \mathbf{k} \right) & 1 \\
      \end{array}
    \right) \, , \label{HSfull}
\end{eqnarray}
where we use the subscript `1' to stress that these matrices
apply to monolayer graphene. The corresponding energy $E$ may be determined
by solving the secular equation $\det \left( H_1 -  E S_1 \right) = 0$,
(\ref{mcc:sec}):
\begin{eqnarray}
\det \! \left(
      \begin{array}{cc}
        \epsilon_{2p} - E & - \left( \gamma_0 + E s_0 \right) f \left( \mathbf{k} \right) \\
        - \left( \gamma_0 + E s_0 \right) f^{\ast} \left( \mathbf{k} \right) & \epsilon_{2p} - E \\
      \end{array}
    \right) &=& 0 \, , \\
\Rightarrow \quad \left( E - \epsilon_{2p} \right)^2
- \left( \left[ E - \epsilon_{2p} \right] s_0 + \epsilon_{2p} s_0 + \gamma_0 \right)^2
|f \left( \mathbf{k} \right) |^2 &=& 0 \, .
\end{eqnarray}
Solving this quadratic equation yields the energy:
\begin{eqnarray}
E_{\pm} = \frac{\epsilon_{2p} \pm \gamma_0 |f \left( \mathbf{k} \right) |}{1 \mp s_0 |f \left( \mathbf{k} \right) |} \, . \label{mcc:monofull}
\end{eqnarray}
This expression appears in Saito {\em et al} \cite{saito}, where
parameter values $\gamma_0 = 3.033\,$eV, $s_0 = 0.129$, $\epsilon_{2p} = 0$
are quoted. The latter value ($\epsilon_{2p} = 0$) means that the zero of
energy is set to be equal to the energy of the $2p_z$ orbital.
The resulting band structure $E_{\pm}$ is shown in Fig.~\ref{mccfig:17} in the vicinity
of the Brillouin zone.
A particular cut through the band structure is shown in Fig.~\ref{mccfig:4} where
the bands are plotted as a function of wave vector component $k_x$ along
the line $k_y = 0$, a line that passes through the center of the
Brillouin zone, labeled $\Gamma$, and two corners of the Brillouin zone, labeled
$K_{+}$ and $K_{-}$ (see the inset of Fig.~\ref{mccfig:4}).
The Fermi level in pristine graphene is located at zero energy.
There are two energy bands, that we refer to as the conduction band ($E_{+}$)
and the valence band ($E_{-}$). The interesting feature of the band structure
is that there is no band gap between the conduction and valence bands. Instead
the bands cross at the six corners of the Brillouin zone, Fig.~\ref{mccfig:17}.
The corners of the Brillouin zone are known as $K$ points,
 and two of them are explicitly labeled $K_{+}$ and $K_{-}$
in Fig.~\ref{mccfig:17}. Near these points, the dispersion is linear
and electronic properties may be described by a Dirac-like Hamiltonian.
This will be explored in more detail in the next section.
Note also that the band structure displays a large asymmetry
between the conduction and valence bands that is most pronounced in the
vicinity of the $\Gamma$ point. This arises from the non-zero overlap
parameter $s_0$ appearing in (\ref{mcc:monofull}).

The tight-binding model described here cannot be used to determine
the values of parameters such as $\gamma_0$ and $s_0$. They must be determined
either by an alternative theoretical method, such as density-functional theory,
or by comparison of the tight-binding model with experiments. Note, however,
that the main qualitative features described in this chapter do not depend on
the precise values of the parameters quoted.

\section{Massless chiral quasiparticles in monolayer graphene}
\label{s:mcc:chiral}

\subsection{The Dirac-like Hamiltonian}
\label{ss:mcc:Dirac}

As described in the previous section, the electronic band structure
of monolayer graphene, Figs.~\ref{mccfig:17}, \ref{mccfig:4}, is gapless, with
crossing of the bands at points $K_{+}$ and $K_{-}$ located
at corners of the Brillouin zone.
In this section, we show that electronic properties near these points may
be described by a Dirac-like Hamiltonian.

Although the first Brillouin zone has six corners,
only two of them are non-equivalent. In this Chapter, we choose
points $K_{+}$ and $K_{-}$, Figs.~\ref{mccfig:17}, \ref{mccfig:4}, as a non-equivalent pair.
It is possible to connect two of the other corners to $K_{+}$ using
a reciprocal lattice vector (hence, the other two are equivalent to $K_{+}$),
and it is possible to connect the remaining two corners to $K_{-}$ using
a reciprocal lattice vector (hence, the remaining two are equivalent to $K_{-}$),
but it is not possible to connect $K_{+}$ and $K_{-}$ with a reciprocal
lattice vector. To distinguish between $K_{+}$ and $K_{-}$, we will use an index
$\xi = \pm 1$. Using the values of the primitive reciprocal
lattice vectors $\mathbf{b}_1$ and $\mathbf{b}_2$, (\ref{mcc:b1b2}),
it can be seen that the wave vector corresponding to point $K_{\xi}$ is given by
\begin{eqnarray}
\mathbf{K}_{\xi} = \xi \left( \frac{4 \pi}{3a} , 0 \right) \, . \label{mcc:kxi}
\end{eqnarray}
Note that the $K$ points are often called `valleys' using nomenclature
from semiconductor physics.

In the tight-binding model, coupling between the $A$ and $B$ sublattices
is described by the off-diagonal matrix element $H_{AB}$, (\ref{mcc:Hoffdiag}), that
is proportional to parameter $\gamma_0$ and the function $f (\mathbf{k})$,
(\ref{mcc:fk0}). Exactly at the $K_{\xi}$ point, $\mathbf{k} = \mathbf{K}_{\xi}$,
the latter is equal to
\begin{eqnarray}
f \left( \mathbf{K}_{\xi} \right)
&=& e^{0} + e^{i \xi 2 \pi /3} + e^{- i \xi 2 \pi /3} = 0 \, . \label{mcc:cancel}
\end{eqnarray}
This indicates that there is no coupling between the $A$ and $B$ sublattices
exactly at the $K_{\xi}$ point. Since the two sublattices are both hexagonal
Bravais lattices of carbon atoms, they support the same quantum states, leading to
a degeneracy point in the spectrum at $K_{\xi}$, Figs.~\ref{mccfig:17}, \ref{mccfig:4}.

The exact cancelation of the three factors describing coupling between
the $A$ and $B$ sublattices, (\ref{mcc:cancel}), no longer holds when the
wave vector is not exactly equal to that of the $K_{\xi}$ point.
We introduce a momentum $\mathbf{p}$ that is measured from the
center of the $K_{\xi}$ point,
\begin{eqnarray}
\mathbf{p} = \hbar \mathbf{k} - \hbar \mathbf{K}_{\xi} \, .
\end{eqnarray}
Then, the coupling between the $A$ and $B$ sublattices is proportional to
\begin{eqnarray}
f \left( \mathbf{k} \right)
&=& e^{i p_y a /\sqrt{3}\hbar} + 2 e^{- i p_y a /2\sqrt{3}\hbar}
\cos \left( \frac{2 \pi \xi}{3} + \frac{p_x a}{2 \hbar} \right) \, , \\
&\approx& \left( 1 + \frac{i p_y a}{\sqrt{3}\hbar} \right) + 2 \left( 1 - \frac{i p_y a}{2\sqrt{3}\hbar} \right)
\left( - \frac{1}{2}
- \frac{\xi\sqrt{3}p_x a}{4\hbar} \right) \, , \\
&\approx& - \frac{\sqrt{3} a}{2 \hbar} \left( \xi p_x - i p_y \right) \, , \label{mcc:flinear}
\end{eqnarray}
where we kept only linear terms in the momentum $\mathbf{p} = \left( p_x , p_y \right)$,
an approximation that is valid close to the $K_{\xi}$ point, {i.e.} for $p a / \hbar \ll 1$,
where $p = |\mathbf{p}| = ( p_x^2 + p_y^2 )^{1/2}$.
Using this approximate expression for the function $f \left( \mathbf{k} \right)$,
the transfer integral matrix (\ref{HSfull}) in the vicinity of point $K_{\xi}$
becomes
\begin{eqnarray}
H_{1,\xi} = v \left(
      \begin{array}{cc}
        0 &   \xi p_x - i p_y  \\
         \xi p_x + i p_y  & 0 \\
      \end{array}
    \right) \, . \label{mcc:dirac1}
\end{eqnarray}
Here, we used $\epsilon_{2p} = 0$ \cite{saito} which defines the
zero of the energy axis to coincide with the energy of the $2p_z$ orbital.
The parameters $a$ and $\gamma_0$ were combined into a velocity $v$ defined as
$v = \sqrt{3} a \gamma_0/ (2 \hbar )$.

Within the linear-in-momentum
approximation for $f \left( \mathbf{k} \right)$, (\ref{mcc:flinear}),
the overlap matrix $S_1$ may be regarded as a unit matrix, because its off-diagonal
elements, proportional to $s_0$, only contribute quadratic-in-momentum
terms to the energy $E_{\pm}$, (\ref{mcc:monofull}).
Since $S_1$ is approximately equal to a unit matrix,
(\ref{mcc:HES}) becomes $H_1 \psi = E \psi$, indicating
that $H_1$, (\ref{mcc:dirac1}), is an effective Hamiltonian for monolayer graphene at low-energy.
The energy eigenvalues and eigenstates of $H_1$ are given by
\begin{eqnarray}
E_{\pm} = \pm v p \, , \qquad
\psi_{\pm} = \frac{1}{\sqrt{2}} \left(
                                  \begin{array}{c}
                                    1 \\
                                    \pm \xi e^{i \xi \varphi} \\
                                  \end{array}
                                \right)
e^{i \mathbf{p} . \mathbf{r} / \hbar} \, , \label{mcc:evalues}
\end{eqnarray}
where $\pm$ refer to the conduction and valence bands, respectively.
Here $\varphi$ is the polar angle of the momentum in the
graphene plane,
$\mathbf{p} = \left( p_x , p_y \right) = p \left( \cos \varphi , \sin \varphi \right)$.

\subsection{Pseudospin and chirality in graphene}
\label{ss:mcc:pseudo}

\begin{figure}[t]
\centering
\includegraphics*[width=0.9\textwidth]{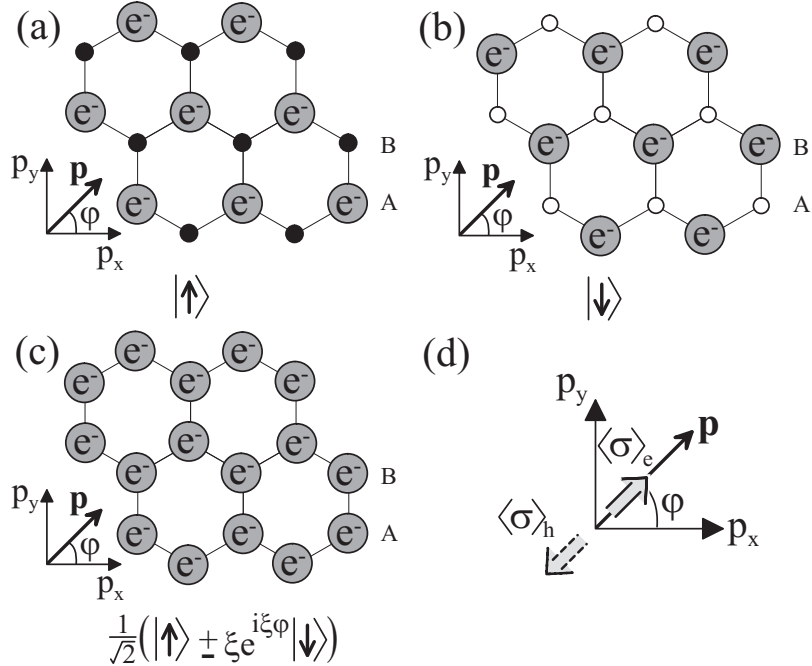}
\caption[]{Schematic representation of the pseudospin degree of freedom:
(a) electronic density solely on the $A$ sublattice can be viewed as a pseudospin
`up' state, whereas (b) density solely on the $B$ sublattice corresponds to a pseudospin
`down' state; (c) in graphene, electronic density is usually shared
equally between $A$ and $B$ sublattices, so that the pseudospin part of
the wave function is a linear combination of `up' and `down,' with amplitudes
dependent on the direction of the electronic momentum $\mathbf{p}$;
(d) at valley $K_{+}$, the pseudospin $\langle \mbox{\boldmath$\sigma$} \rangle_{e}$
in the conduction band is parallel to the momentum,
whereas the pseudospin $\langle \mbox{\boldmath$\sigma$} \rangle_{h}$
in the valence band is anti-parallel to the momentum.}
\label{mccfig:5}
\end{figure}

The effective Hamiltonian (\ref{mcc:dirac1}) and eigenstates
(\ref{mcc:evalues}) in the vicinity of the $K_{\xi}$ point
have two components, reminiscent of the components of spin-$1/2$.
Referring back to the original definitions of the components
of the column vector $\psi$, (\ref{mcc:exp}) and (\ref{mcc:HSpsi}),
shows that this is not the physical spin of the electron, but
a degree of freedom related to the relative amplitude of the Bloch function
on the $A$ or $B$ sublattice. This degree of freedom is called pseudospin.
If all the electronic density was located on the $A$ sublattice,
Fig.~\ref{mccfig:5}(a), this could be viewed as a pseudospin
`up' state (pointing upwards out of the graphene sheet)
$| \!\!\uparrow \rangle = ( 1 , 0 )^{T}$,
whereas density solely on the $B$ sublattice corresponds to a pseudospin
`down' state (pointing downwards out of the graphene sheet)
$| \!\!\downarrow \rangle = ( 0 , 1 )^{T}$, Fig.~\ref{mccfig:5}(b).
In graphene, electronic density is usually shared
equally between $A$ and $B$ sublattices, Fig.~\ref{mccfig:5}(c),
so that the pseudospin part of the wave function is a linear combination
of `up' and `down,' and it lies in the plane of the graphene sheet.

Not only do the electrons possess the pseudospin degree of freedom,
but they are chiral, meaning that the orientation of the pseudospin
is related to the direction of the electronic momentum $\mathbf{p}$.
This is reflected in the fact that the amplitudes on the $A$ or $B$ sublattice
of the eigenstate (\ref{mcc:evalues}) depend on the polar angle
$\varphi$.
It is convenient to use Pauli spin matrices in the
$A$/$B$ sublattice space, $\sigma_i$ where $i = 1 \ldots 3$, to write
the effective Hamiltonian (\ref{mcc:dirac1}) as
\begin{eqnarray}
H_{1,\xi} = v \left( \xi \sigma_x p_x + \sigma_y p_y \right) \, . \label{mcc:dirac2}
\end{eqnarray}
If we define a pseudospin vector as
$\mbox{\boldmath$\sigma$} = \left( \sigma_x , \sigma_y , \sigma_z \right)$,
and a unit vector as
$\mathbf{\hat n}_1 = \left( \xi \cos \varphi , \sin \varphi , 0 \right)$,
then the Hamiltonian becomes
$H_{1,\xi} = v p \,\mbox{\boldmath$\sigma$}.\mathbf{\hat n}_1$,
stressing that the pseudospin $\mbox{\boldmath$\sigma$}$ is linked to the
direction $\mathbf{\hat n}_1$.
The chiral operator $\mbox{\boldmath$\sigma$}.\mathbf{\hat n}_1$ projects
the pseudospin onto the direction of quantization $\mathbf{\hat n}_1$:
eigenstates of the Hamiltonian are also eigenstates of
$\mbox{\boldmath$\sigma$}.\mathbf{\hat n}_1$ with eigenvalues $\pm 1$,
$\mbox{\boldmath$\sigma$}.\mathbf{\hat n}_1 \psi_{\pm} = \pm \psi_{\pm}$.
An alternative way of expressing this chiral property of
electrons is to explicitly calculate the
expectation value of the pseudospin operator
$\langle \mbox{\boldmath$\sigma$} \rangle
= \left( \langle \sigma_x \rangle , \langle \sigma_y \rangle , \langle \sigma_z \rangle \right)$
with respect to the eigenstate $\psi_{\pm}$, (\ref{mcc:evalues}). The result,
$\langle \mbox{\boldmath$\sigma$} \rangle_{e/h}
= \pm \left( \xi \cos \varphi , \sin \varphi , 0 \right)$,
shows the link between pseudospin and momentum.
For valley $K_{+}$, the pseudospin in the conduction band
$\langle \mbox{\boldmath$\sigma$} \rangle_{e}$ is parallel to the
momentum, whereas the pseudospin in the valence band
$\langle \mbox{\boldmath$\sigma$} \rangle_{h}$ is anti-parallel
to it, Fig.~\ref{mccfig:5}(d).

\begin{figure}[t]
\centering
\includegraphics*[width=1.0\textwidth]{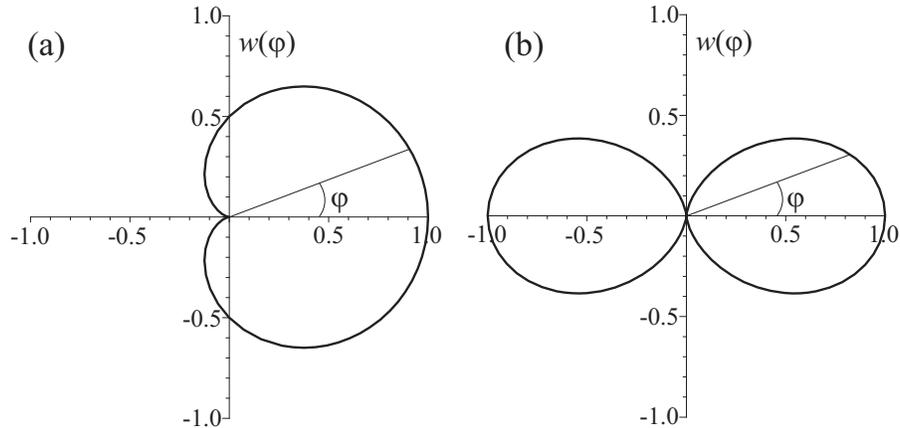}
\caption[]{Anisotropic scattering of chiral electrons in graphene:
(a) angular dependence $w(\varphi) = \cos^2(\varphi/2)$ of the scattering probability
off an $A$-$B$ symmetric potential in monolayer graphene \cite{ando98,mceuen99,suz02} and
(b) $w(\varphi) = \cos^2(\varphi)$ in bilayer graphene \cite{mcc06a,falko07}.}
\label{mccfig:13}
\end{figure}

If the electronic momentum $\mathbf{p}$ rotates by angle $\varphi$,
then adiabatic evolution of the chiral wave function $\psi_{\pm}$, (\ref{mcc:evalues}),
produces a matching rotation of the vector $\mathbf{\hat n}_1$ by angle $\varphi$.
For traversal of a closed contour in momentum space, corresponding to $\varphi = 2\pi$,
then the chiral wave function undergoes a phase change of $\pi$ known as
Berry's phase \cite{pan56,berry84}. It can be thought of as arising from the
rotation of the pseudospin degree of freedom.

The chiral nature of low-energy electrons in graphene places an additional constraint
on their scattering properties. If a given potential doesn't break the
$A$-$B$ symmetry, then it is unable to influence the pseudospin degree
of freedom which must, therefore, be conserved upon scattering.
Considering only the pseudospin part of the chiral wave function $\psi_{\pm}$,
(\ref{mcc:evalues}), the probability to scatter in a direction $\varphi$,
where $\varphi = 0$ is the forwards direction, is proportional to
$w (\varphi) = |\langle \psi_{\pm} (\varphi) | \psi_{\pm} (0) \rangle |^2$.
For monolayer graphene, $w (\varphi) = \cos^2 (\varphi/2)$, Fig.~\ref{mccfig:13}(a).
This is anisotropic, and displays an absence of backscattering
$w (\pi) = 0$ \cite{ando98,mceuen99,suz02}: scattering into a state with
opposite momentum is prohibited because it
requires a reversal of the pseudospin. Such conservation of pseudospin
is at the heart of anisotropic scattering at
potential barriers in graphene monolayers \cite{che06,kat06},
known as Klein tunneling.

\section{The tight-binding model of bilayer graphene}
\label{s:mcc:tbmbilayer}

In this section, we describe the tight-binding model of bilayer graphene.
To do so, we use the tight-binding model described in
Sect.~\ref{s:mcc:tbmgeneral} in order to generalize the model
for monolayer graphene discussed in Sect.~\ref{s:mcc:tbmmonolayer}.

\begin{figure}[t]
\centering
\includegraphics*[width=0.6\textwidth]{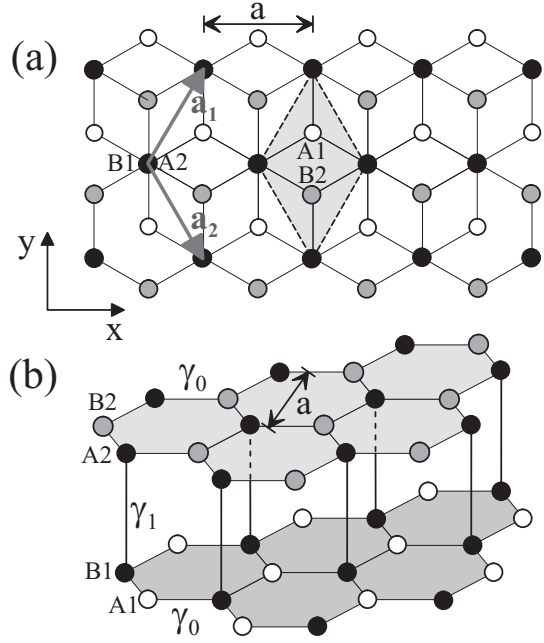}
\caption[]{Schematic representation of the crystal structure of
$AB$-stacked bilayer graphene:
(a) plan view with $A1$ (white) and $B1$ atoms (black) on the
lower layer, $A2$ (black) and $B2$ atoms (grey) on the
upper layer. Vectors $\mathbf{a}_1$ and $\mathbf{a}_2$ are primitive lattice
vectors of length equal to the lattice constant $a$, and
the shaded rhombus is a unit cell; (b) side view where the parameter $\gamma_0$
represents nearest-neighbor coupling within each layer,
$\gamma_1$ nearest-neighbor coupling between the $B1$ and $A2$ atoms
on different layers.}
\label{mccfig:6}
\end{figure}

We consider Bernal-stacked bilayer graphene \cite{trickey,novo06,mcc06a}
(also called $AB$-stacked bilayer graphene).
It consists of two parallel layers of carbon atoms, each arranged with a honeycomb
arrangement as in a monolayer, that are coupled together, Fig.~\ref{mccfig:5}. There are
four atoms in the unit cell, a pair $A1$, $B1$, from the lower layer
and a pair $A2$, $B2$, from the upper layer.
In Bernal stacking, the layers are arranged so that two atoms,
$B1$ and $A2$, are directly below or above each other, whereas the
other two atoms, $A1$ and $B2$, do not have a counterpart in the other layer.
The primitive lattice vectors $\mathbf{a}_1$ and $\mathbf{a}_2$,
and the lattice constant $a$ are the same as for monolayer graphene,
and the unit cell, shown in Fig.~\ref{mccfig:5}(a), has the same
area in the $x$-$y$ plane as in the monolayer. Therefore, the reciprocal lattice and first
Brillouin zone are the same as in monolayer graphene, Fig.~\ref{mccfig:2}.
The unit cell of bilayer graphene contains
four atoms, and, if the tight-binding model includes one $p_z$ orbital
per atomic site, there will be four bands near zero energy, instead
of the two bands in monolayer graphene.

Essential features of the low-energy electronic band structure may
be described by a minimal tight-binding model including nearest-neighbor
coupling $\gamma_0$ between $A1$ and $B1$, and $A2$ and $B2$, atoms
on each layer, and nearest-neighbor interlayer coupling $\gamma_1$
between $B1$ and $A2$ atoms that are directly below or above each other,
\begin{eqnarray}
\gamma_1 =
\langle \phi_{A2} \left( \mathbf{r} - \mathbf{R}_{A2} \right) | {\cal H} | \phi_{B1} \left( \mathbf{r} - \mathbf{R}_{B1} \right) \rangle \, .
\end{eqnarray}
Then, we can generalize the treatment of monolayer
graphene, (\ref{HSfull}), to write the transfer and overlap integral matrices of bilayer
graphene, in a basis with components $A1$, $B1$, $A2$, $B2$, as
\begin{eqnarray}
H &=& \left(
      \begin{array}{cccc}
        \epsilon_{2p} & - \gamma_0 f \left( \mathbf{k} \right) & 0 & 0 \\
        - \gamma_0 f^{\ast} \left( \mathbf{k} \right) & \epsilon_{2p} & \gamma_1 & 0 \\
        0 & \gamma_1 & \epsilon_{2p} & - \gamma_0 f \left( \mathbf{k} \right) \\
        0 & 0 & - \gamma_0 f^{\ast} \left( \mathbf{k} \right) & \epsilon_{2p} \\
      \end{array}
    \right) \,  ,  \label{mcc:Hbifull} \\
S &=& \left(
      \begin{array}{cccc}
        1 & s_0 f \left( \mathbf{k} \right) & 0 & 0 \\
        s_0 f^{\ast} \left( \mathbf{k} \right) & 1 & 0 & 0 \\
        0 & 0 & 1 & s_0 f \left( \mathbf{k} \right) \\
        0 & 0 & s_0 f^{\ast} \left( \mathbf{k} \right) & 1 \\
      \end{array}
    \right) \,  . \label{mcc:Sbifull}
\end{eqnarray}
The upper-left and lower-right $2 \times 2$ blocks describe behavior within
the lower ($A1$/$B1$) and upper ($A2$/$B2$) layers, respectively. The off-diagonal
$2 \times 2$ blocks, containing parameter $\gamma_1$, describe interlayer coupling.

\begin{figure}[t]
\centering
\includegraphics*[width=.7\textwidth]{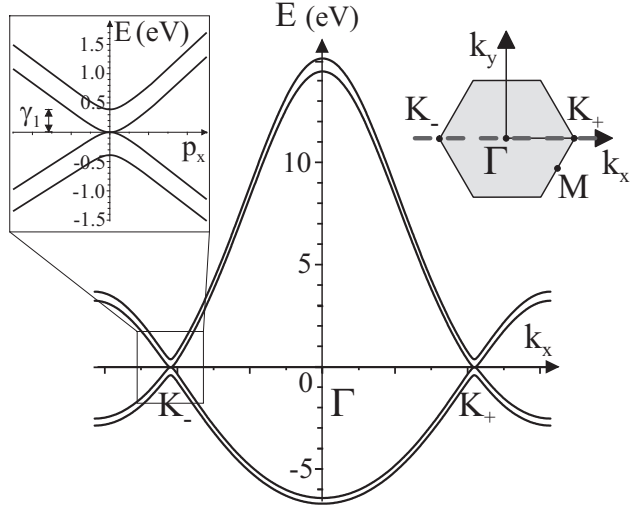}
\caption[]{The low-energy band structure of bilayer
graphene taking into account nearest-neighbor hopping
with parameter $\gamma_0 = 3.033$eV, nearest-neighbor overlap
parameter $s_0 = 0.129$, orbital energy $\epsilon_{2p} = 0$ \cite{saito},
and interlayer coupling $\gamma_1 = 0.39$eV.
The plot shows the bands calculated along the $k_x$ axis intersecting
points $K_{-}$, $\Gamma$, and $K_{+}$ in the Brillouin zone,
shown as the dotted line in the right inset. The left inset shows the
band structure in the vicinity of the point $K_{-}$.}
\label{mccfig:7}
\end{figure}

The band structure of bilayer graphene may be determined by solving the secular equation
$\det \left( H -  E_j S \right) = 0$, (\ref{mcc:sec}).
It is plotted in Fig.~\ref{mccfig:7} for parameter values
$\gamma_0 = 3.033$eV, $s_0 = 0.129$, $\epsilon_{2p} = 0$ \cite{saito}
and interlayer coupling $\gamma_1 = 0.39$eV.
There are four energy bands, two conduction bands and two valence bands.
Overall, the band structure is similar to that of monolayer graphene,
Fig.~\ref{mccfig:4}, with each monolayer band split into two by an energy
approximately equal to the interlayer coupling $\gamma_1$ \cite{trickey}.
The most interesting part of the band structure is in the vicinity of
the $K$ points \cite{mcc06a}, as shown in the left inset of Fig.~\ref{mccfig:7}
which focuses in on the bands around $K_{-}$. At the $K$ point,
one of the conduction (valence) bands is split away from zero energy
by an amount equal to the interlayer coupling $\gamma_1$ (-$\gamma_1$).
The split bands originate from atomic sites $B1$ and $A2$
that have a counterpart atom directly above or below
them on the other layer. Orbitals on these pairs of atoms ($B1$ and $A2$) are strongly coupled
by the interlayer coupling $\gamma_1$ and they form a bonding and anti-bonding
pair of bands, split away from zero energy. In the following,
we refer to them as `dimer' states, and atomic sites $B1$ and $A2$
are called `dimer' sites.
The remaining two bands, one conduction and one valence band, touch
at zero energy: as in the monolayer, there is no
band gap between the conduction and valence bands.
In the vicinity of the $K$ points, the dispersion of the latter bands
is quadratic $E_{\pm} \propto \pm |\mathbf{k} - \mathbf{K}_{\xi}|^2$,
and electronic properties of the low-energy bands may be described by an
effective Hamiltonian describing massive chiral particles.
This will be explored in more detail in the next section.

\section{Massive chiral quasiparticles in bilayer graphene}
\label{s:mcc:massive}

\begin{figure}[t]
\centering
\includegraphics*[width=0.5\textwidth]{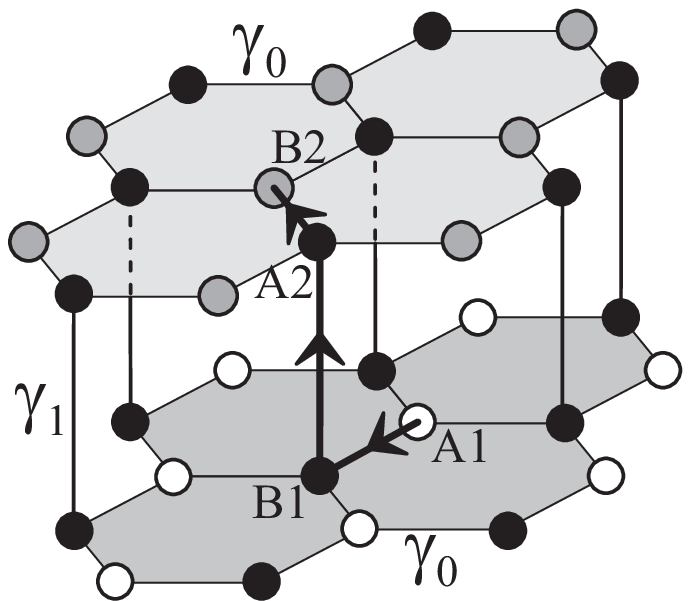}
\caption[]{Schematic representation of the crystal structure of
$AB$-stacked bilayer graphene illustrating the processes that contribute
to effective coupling between $A1$ (white) and $B2$ atoms (grey),
in the presence of strongly-coupled `dimer' sites $B1$ and $A2$ (black).
The black arrowed line indicates the three stage process:
intralayer hopping between $A1$ and $B1$, followed
by an interlayer transition via the dimer sites $B1$ and $A2$,
followed by another intralayer hopping between $A2$ and $B2$.}
\label{mccfig:8}
\end{figure}

\subsection{The low-energy bands of bilayer graphene}
\label{ss:mcc:bilow}

To begin the description of the low-energy bands in bilayer graphene,
we set $s_0=0$, thus neglecting the non-orthogonality of orbitals
that tends to become important at high energy. Then, the
overlap matrix $S$, (\ref{mcc:Sbifull}), becomes a unit matrix,
and $H$, (\ref{mcc:Hbifull}), is an effective Hamiltonian for the four
bands of bilayer graphene at low-energy \cite{mcc06a}:
\begin{eqnarray}
H = \left(
      \begin{array}{cccc}
        0 & - \gamma_0 f \left( \mathbf{k} \right) & 0 & 0 \\
        - \gamma_0 f^{\ast} \left( \mathbf{k} \right) & 0 & \gamma_1 & 0 \\
        0 & \gamma_1 & 0 & - \gamma_0 f \left( \mathbf{k} \right) \\
        0 & 0 & - \gamma_0 f^{\ast} \left( \mathbf{k} \right) & 0 \\
      \end{array}
    \right) \,  , \label{mcc:Hbifull2}
\end{eqnarray}
where we used $\epsilon_{2p} = 0$ \cite{saito} to define the
zero of the energy axis to coincide with the energy of the $2p_z$ orbital.
Eigenvalues of the Hamiltonian are given by
\begin{eqnarray}
E_{\pm}^{(\alpha)} = \pm \frac{\gamma_1}{2}
\left( \sqrt{1 + \frac{4 \gamma_0^2 |f(\mathbf{k})|^2}{\gamma_1^2}} + \alpha \right) \, ,
\qquad \alpha = \pm 1 \, .
\label{mcc:Efour}
\end{eqnarray}
Over most of the Brillouin zone, where $4 \gamma_0^2 |f(\mathbf{k})|^2 \gg \gamma_1^2$,
the energy may be approximated as
$E_{\pm}^{(\alpha)} \approx \pm ( \gamma_0 |f(\mathbf{k})| + \alpha \gamma_1 / 2)$,
meaning that the $\alpha = \pm 1$ bands are approximately the same as the monolayer
bands, (\ref{mcc:monofull}), but they are split by the interlayer coupling $\gamma_1$.
The eigenvalues $E_{\pm}^{(1)}$, (\ref{mcc:Efour}),
describe two bands that are split away from zero energy by $\pm \gamma_1$
at the $K$ point (where $|f(\mathbf{k})| = 0$) as shown in the
left inset of Fig.~\ref{mccfig:7}.
This is because the orbitals on the $A2$ and $B1$ sites form a dimer
that is coupled by interlayer hopping $\gamma_1$, resulting in a bonding and
anti-bonding pair of states $\pm \gamma_1$.

The remaining two bands are described by $E_{\pm}^{(-1)}$.
Near to the $K_{\xi}$ point, $p a / \hbar \ll 1$, we replace the
factor $\gamma_0 |f(\mathbf{k})|$ with $vp$, (\ref{mcc:flinear}):
\begin{eqnarray}
E_{\pm}^{(-1)} \approx \pm \frac{\gamma_1}{2}
\left( \sqrt{1 + \frac{4 v^2 p^2}{\gamma_1^2}} - 1 \right)  \, .
\label{mcc:Efourlow}
\end{eqnarray}
This formula interpolates between linear dispersion at large momenta
($\gamma_1 \ll vp < \gamma_0$) and
quadratic dispersion
$E_{\pm}^{(-1)} \approx \pm v^2p^2 / \gamma_1$
near zero energy where the bands touch.
These bands arise from effective coupling between the orbitals on
sites, $A1$ and $B2$, that don't have a counterpart in the other layer.
In the absence of direct coupling between $A1$ and $B2$, the effective coupling
is achieved through a three stage process as indicated in Fig.~\ref{mccfig:8}.
It can be viewed as an intralayer hopping between $A1$ and $B1$, followed
by an interlayer transition via the dimer sites $B1$ and $A2$,
followed by another intralayer hopping between $A2$ and $B2$.
This effective coupling may be succinctly described by an effective low-energy
Hamiltonian written in a two-component basis of $p_z$ orbitals on $A1$ and $B2$
sites.

\subsection{The two-component Hamiltonian of bilayer graphene}
\label{ss:mcc:bitwo}

The effective two-component Hamiltonian may be derived from
the four component Hamiltonian, (\ref{mcc:Hbifull2}),
using a Schrieffer-Wolff transformation \cite{swtrans,mcc06a}.
In the present context, a straightforward way to do the transformation
is to consider the eigenvalue equation for
the four component Hamiltonian, (\ref{mcc:Hbifull2}),
as four simultaneous equations for the wave-function
components ${c}_{A1}$, ${c}_{B1}$, ${c}_{A2}$, ${c}_{B2}$:
\begin{eqnarray}
E{c}_{A1} + \gamma_0 f \left( \mathbf{k} \right) {c}_{B1} &=& 0 \, , \label{mcc:se1} \\
\gamma_0 f^{\ast} \left( \mathbf{k} \right){c}_{A1} + E {c}_{B1} - \gamma_1 {c}_{A2} &=& 0 \, , \label{mcc:se2} \\
- \gamma_1 {c}_{B1} + E {c}_{A2} + \gamma_0 f \left( \mathbf{k} \right) {c}_{B2} &=& 0 \, , \label{mcc:se3} \\
\gamma_0 f^{\ast} \left( \mathbf{k} \right){c}_{A2} + E {c}_{B2} &=& 0 \, \label{mcc:se4} .
\end{eqnarray}
Using the second and third equations, (\ref{mcc:se2}) and (\ref{mcc:se3}), it is possible to express
the components on the dimer sites, ${c}_{B1}$ and ${c}_{A2}$, in
terms of the other two:
\begin{eqnarray}
{c}_{B1} &=& \frac{\gamma_0 f \left( \mathbf{k} \right)}{\gamma_1 d} {c}_{B2}
+ \frac{E \gamma_0 f^{\ast} \left( \mathbf{k} \right)}{\gamma_1^2 d} {c}_{A1} , \\
{c}_{A2} &=& \frac{E \gamma_0 f \left( \mathbf{k} \right)}{\gamma_1^2 d} {c}_{B2}
+ \frac{\gamma_0 f^{\ast} \left( \mathbf{k} \right)}{\gamma_1 d} {c}_{A1} \, ,
\end{eqnarray}
where $d = 1 - E^2/\gamma_1^2$.
Substituting these expressions into the first and fourth equations,
(\ref{mcc:se1}) and (\ref{mcc:se4}), produces
two equations solely in terms of ${c}_{A1}$ and ${c}_{B2}$.
Assuming $|E| \ll |\gamma_1|$ and $|\gamma_0 f \left( \mathbf{k} \right)| \ll |\gamma_1|$,
we use $d \approx 1$ and keep terms up to order $1/\gamma_1$ only:
\begin{eqnarray}
E{c}_{A1}
+ \frac{\gamma_0^2 f^2 \left( \mathbf{k} \right)}{\gamma_1} {c}_{B2} &=& 0 \, , \\
\frac{\gamma_0^2 (f^{\ast}\left( \mathbf{k} \right))^2}{\gamma_1} {c}_{A1}
+ E {c}_{B2} &=& 0 \, .
\end{eqnarray}
It is possible to express these two equations
as a Schr\"odinger equation, $H_2 \psi = E \psi$, with a two-component
wave function $\psi = \left( {c}_{A1} , {c}_{B2} \right)^{T}$ and
two-component Hamiltonian
\begin{eqnarray}
H_{2,\xi} = - \frac{1}{2m} \left(
      \begin{array}{cc}
        0 & \left( \xi p_x - i p_y \right)^2 \\
        \left( \xi p_x + i p_y \right)^2 & 0 \\
      \end{array}
    \right) \, , \label{mcc:bi1}
\end{eqnarray}
where we used the approximation
$f \left( \mathbf{k} \right) \approx - v \left( \xi p_x - i p_y \right)/\gamma_0$,
(\ref{mcc:flinear}), valid for momentum $p a / \hbar \ll 1$ close to the $K_{\xi}$ point,
and parameters $v$ and $\gamma_1$ were combined into a mass $m = \gamma_1 / (2v^2)$.

The effective low-energy Hamiltonian of bilayer graphene,
(\ref{mcc:bi1}),  resembles the Dirac-like Hamiltonian of monolayer
graphene, (\ref{mcc:dirac1}), but with a quadratic term on
the off-diagonal instead of linear.
The energy eigenvalues and eigenstates of $H_2$ are given by
\begin{eqnarray}
E_{\pm} = \pm \frac{p^2}{2m} \, , \qquad
\psi_{\pm} = \frac{1}{\sqrt{2}} \left(
                                  \begin{array}{c}
                                    1 \\
                                    \mp e^{i 2 \xi \varphi} \\
                                  \end{array}
                                \right)
e^{i \mathbf{p} . \mathbf{r} / \hbar} \, , \label{mcc:bievalues}
\end{eqnarray}
where $\pm$ refer to the conduction and valence bands, respectively.
Here $\varphi$ is the polar angle of the momentum in the
graphene plane,
$\mathbf{p} = \left( p_x , p_y \right) = p \left( \cos \varphi , \sin \varphi \right)$.

\subsection{Pseudospin and chirality in bilayer graphene}
\label{ss:mcc:bipseudo}

\begin{figure}[t]
\centering
\includegraphics*[width=0.8\textwidth]{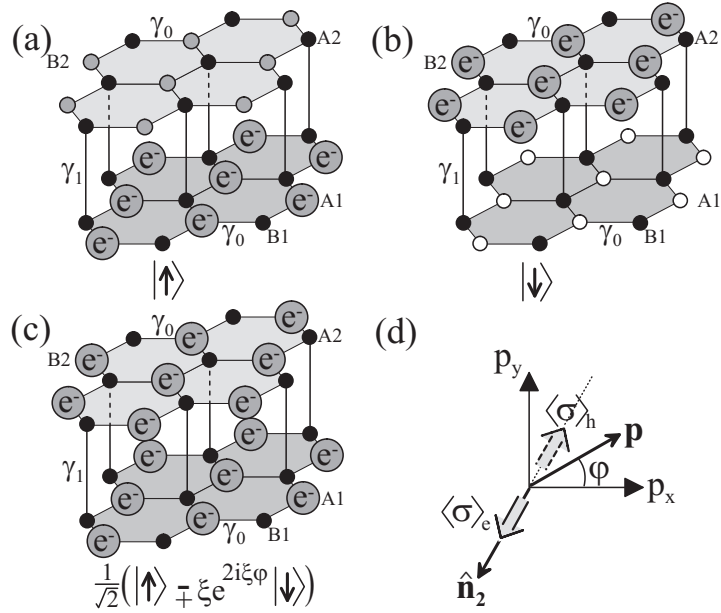}
\caption[]{Schematic representation of the pseudospin degree of freedom
in bilayer graphene:
(a) electronic density solely on the $A1$ sublattice on the lower
layer can be viewed as a pseudospin `up' state, whereas (b) density
solely on the $B2$ sublattice on the upper layer corresponds to a pseudospin
`down' state; (c) in bilayer graphene, electronic density is usually shared
equally between $A1$ and $B2$ sublattices, so that the pseudospin part of
the wave function is a linear combination of `up' and `down,' with amplitudes
dependent on the direction of the electronic momentum $\mathbf{p}$;
(d) at valley $K_{+}$, the pseudospin $\langle \mbox{\boldmath$\sigma$} \rangle_{e}$
in the conduction band is parallel to the quantization direction $\mathbf{\hat n}_2$,
whereas the pseudospin $\langle \mbox{\boldmath$\sigma$} \rangle_{h}$
in the valence band is anti-parallel to $\mathbf{\hat n}_2$.
Direction $\mathbf{\hat n}_2$ is related to the direction of momentum $\mathbf{p}$,
but turns in the $x$-$y$ plane twice as quickly as it.}
\label{mccfig:9}
\end{figure}

The two-component Hamiltonian (\ref{mcc:bi1}) of bilayer graphene
has a pseudospin degree of freedom \cite{novo06,mcc06a} related to the amplitude of
the eigenstates (\ref{mcc:bievalues}) on the $A1$ and $B2$ sublattice
sites, where $A1$ and $B2$ lie on different layers.
If all the electronic density was located on the $A1$ sublattice,
Fig.~\ref{mccfig:9}(a), this could be viewed as a pseudospin
`up' state (pointing upwards out of the graphene sheet)
$| \!\!\uparrow \rangle = ( 1 , 0 )^{T}$,
whereas density solely on the $B2$ sublattice corresponds to a pseudospin
`down' state (pointing downwards out of the graphene sheet)
$| \!\!\downarrow \rangle = ( 0 , 1 )^{T}$, Fig.~\ref{mccfig:9}(b).
In bilayer graphene, electronic density is usually shared
equally between the two sublattices, Fig.~\ref{mccfig:9}(c),
so that the pseudospin part of the wave function is a linear combination
of `up' and `down,' and it lies in the plane of the graphene sheet.

Electrons in bilayer graphene are chiral \cite{novo06,mcc06a},
meaning that the orientation of the pseudospin
is related to the direction of the electronic momentum $\mathbf{p}$,
but the chirality is different to that in monolayers.
As before, we use Pauli spin matrices in the
$A1$/$B2$ sublattice space, $\sigma_i$ where $i = 1 \ldots 3$, to write
the effective Hamiltonian (\ref{mcc:bi1}) as
\begin{eqnarray}
H_{2,\xi} = - \frac{1}{2m} \left[ \sigma_x \left(p_x^2-p_y^2\right) +
2\xi \sigma_y p_x p_y \right] \, . \label{mcc:bi2}
\end{eqnarray}
If we define a pseudospin vector as
$\mbox{\boldmath$\sigma$} = \left( \sigma_x , \sigma_y , \sigma_z \right)$,
and a unit vector as
$\mathbf{\hat n}_2 = - \left( \cos 2\varphi , \xi \sin 2\varphi , 0 \right)$,
then the Hamiltonian becomes
$H_{2,\xi} = (p^2/2m) \,\mbox{\boldmath$\sigma$}.\mathbf{\hat n}_2$,
stressing that the pseudospin $\mbox{\boldmath$\sigma$}$ is linked to the
direction $\mathbf{\hat n}_2$.
The chiral operator $\mbox{\boldmath$\sigma$}.\mathbf{\hat n}_2$ projects
the pseudospin onto the direction of quantization $\mathbf{\hat n}_2$:
eigenstates of the Hamiltonian are also eigenstates of
$\mbox{\boldmath$\sigma$}.\mathbf{\hat n}_2$ with eigenvalues $\pm 1$,
$\mbox{\boldmath$\sigma$}.\mathbf{\hat n}_2 \psi_{\pm} = \pm \psi_{\pm}$.
In bilayer graphene, the quantization axis $\mathbf{\hat n}_2$ is fixed
to lie in the graphene plane, but it turns twice as quickly in
the plane as the momentum $\mathbf{p}$.
If we calculate the expectation value of the pseudospin operator
$\langle \mbox{\boldmath$\sigma$} \rangle
= \left( \langle \sigma_x \rangle , \langle \sigma_y \rangle , \langle \sigma_z \rangle \right)$
with respect to the eigenstate $\psi_{\pm}$, (\ref{mcc:bievalues}),
then the result $\langle \mbox{\boldmath$\sigma$} \rangle_{e/h}
= \mp \left( \cos 2\varphi , \xi \sin 2\varphi , 0 \right)$,
illustrates the link between pseudospin and momentum, Fig.~\ref{mccfig:9}(d).

If the momentum $\mathbf{p}$ rotates by angle $\varphi$,
adiabatic evolution of the chiral wave function $\psi_{\pm}$, (\ref{mcc:bievalues}),
produces a matching rotation of the quantization axis $\mathbf{\hat n}_2$ by angle $2\varphi$,
not $\varphi$ as in the monolayer, Sect.~\ref{ss:mcc:pseudo}.
Thus, traversal around a closed contour in momentum space results in
a Berry's phase \cite{pan56,berry84} change of $2\pi$ of the chiral
wave function in bilayer graphene \cite{novo06,mcc06a}.
For Berry's phase $2\pi$ chiral electrons in bilayer graphene, (\ref{mcc:bievalues}),
the probability to scatter in a direction $\varphi$,
where $\varphi = 0$ is the forwards direction, is proportional to
$w (\varphi) = |\langle \psi_{\pm} (\varphi) | \psi_{\pm} (0) \rangle |^2
= \cos^2 (\varphi)$ \cite{mcc06a,falko07} as shown in Fig.~\ref{mccfig:13}(b).
This is anisotropic, but, unlike monolayers Fig.~\ref{mccfig:13}(a),
does not display an absence of backscattering
($w (\pi) = 1$ in bilayers): scattering into a state with
opposite momentum is not prohibited because it doesn't
require a reversal of the pseudospin.

\section{The integer quantum Hall effect in graphene}
\label{s:mcc:qhe}

When a perpendicular magnetic field is applied a two-dimensional
electron gas, the electrons follow cyclotron orbits, and their
allowed energies are quantized into values known as Landau levels \cite{landau}.
At low magnetic field, the Landau levels give rise to quantum oscillations
including the de Haas-van Alphen effect and the Shubnikov-de Haas effect.
At higher fields, the discrete Landau level spectrum is manifest in the integer
quantum Hall effect \cite{vk80,p+r87,macdonald89}, a quantization
of Hall conductivity into integer values of the quantum of conductivity $e^2/h$.
For monolayer graphene, the Landau level spectrum was calculated
over fifty years ago by McClure \cite{mcclure56}, and the integer quantum
Hall effect was observed \cite{novo05,zhang05} and studied
theoretically \cite{haldane88,zheng02,gusynin05,peres06,herbut07} in recent years.
The chiral nature of electrons in graphene results in an unusual
sequencing of the quantized plateaus of the Hall conductivity.
In bilayer graphene, the experimental observation of the integer quantum
Hall effect \cite{novo06} and calculation of the Landau level spectrum \cite{mcc06a}
revealed further unusual features related to the chirality of electrons.

\subsection{The Landau level spectrum of monolayer graphene}
\label{ss:mcc:llmono}

We consider a magnetic field perpendicular to the graphene sheet
$\mathbf{B} = \left( 0 , 0, - B \right)$ where $B = |\mathbf{B}|$.
The Dirac-like Hamiltonian of monolayer graphene (\ref{mcc:dirac1})
may be written as
\begin{eqnarray}
\!\!\! \!\!\! H_{1,K_{+}} = v \left(
      \begin{array}{cc}
        0 &   \pi^{\dagger}  \\
         \pi  & 0 \\
      \end{array}
    \right) ,
\quad
H_{1,K_{-}} = - v \left(
      \begin{array}{cc}
        0 &   \pi  \\
         \pi^{\dagger}  & 0 \\
      \end{array}
    \right) , \quad
\Big\{\begin{array}{c}
  \pi = p_x + i p_y \\
  \pi^{\dagger} = p_x - i p_y
\end{array} ,
\label{mcc:diracmf}
\end{eqnarray}
in the vicinity of corners of the Brillouin zone
$K_{+}$ and $K_{-}$, respectively.
The off-diagonal elements of the Hamiltonian (\ref{mcc:diracmf}) contain operators
$\pi = p_x + i p_y$ and $\pi^{\dagger} = p_x - i p_y$,
where, in the presence of a magnetic field,
the operator $\mathbf{p} = (p_x , p_y) \equiv -i\hbar \nabla + e \mathbf{A}$.
Here $\mathbf{A}$ is the vector potential
and the charge of the electron is $- e$.

Using the Landau gauge $\mathbf{A} = \left( 0, -Bx, 0 \right)$ preserves
translational invariance in the $y$ direction,
so that eigenstates
may be written in terms of states that are plane waves in the $y$ direction
and harmonic oscillator states in the $x$ direction \cite{p+r87,macdonald89},
\begin{eqnarray}
\phi_{\ell} \left( x , y \right) =
A_{\ell} {\cal H}_{\ell} \!\! \left( \frac{x}{\lambda_B} - \frac{p_y \lambda_B}{\hbar} \right)
\!\exp \!\! \left[ - \frac{1}{2}\left( \frac{x}{\lambda_B} - \frac{p_y \lambda_B}{\hbar} \right)^2
+ i \frac{p_y y}{\hbar} \right] \!\! . \label{mcc:ho}
\end{eqnarray}
Here, ${\cal H}_{\ell}$ are Hermite polynomials of order ${\ell}$, for integer
${\ell} \geq 0$, and
the normalization constant is $A_{\ell} = 1/\sqrt{2^{\ell} {\ell}! \sqrt{\pi}}$.
The magnetic length $\lambda_B$, and a related
energy scale $\Gamma$, are defined as
\begin{eqnarray}
\lambda_B = \sqrt{\frac{\hbar}{eB}} \, , \qquad
\Gamma = \frac{\sqrt{2} \hbar v}{\lambda_B} = \sqrt{2\hbar v^2 eB} \, .
\label{mcc:lb}
\end{eqnarray}
With this choice of vector potential,
$\pi = -i \hbar \partial_x + \hbar \partial_y - i e B x$
and
$\pi^{\dagger} = -i \hbar \partial_x - \hbar \partial_y + i e B x$.
Acting on the harmonic oscillator states (\ref{mcc:ho}) gives
\begin{eqnarray}
\pi \phi_{\ell} &=& - \frac{\sqrt{2} i \hbar}{\lambda_B} \sqrt{{\ell}} \, \phi_{{\ell}-1} \, ,
\label{mcc:lower} \\
\pi^{\dagger} \phi_{\ell} &=& \frac{\sqrt{2} i \hbar}{\lambda_B} \sqrt{{\ell}+1} \, \phi_{{\ell}+1} \, ,
\label{mcc:raise}
\end{eqnarray}
and $\pi \phi_{0}=0$. These equations indicate that operators $\pi$
and $\pi^{\dagger}$ are proportional to lowering and raising operators
of the harmonic oscillator states $\phi_{\ell}$. The Landau level spectrum
is, therefore, straightforward to calculate \cite{mcclure56,fisch70,zheng02}.
At the first valley, $K_{+}$, the Landau level energies and eigenstates
of $H_{1,K_{+}}$ are
\begin{eqnarray}
K_{+}, \, {\ell} \geq 1 : \quad E_{{\ell},\pm} &=& \pm \frac{\sqrt{2} \hbar v}{\lambda_B} \sqrt{{\ell}}  \, , \qquad
\psi_{{\ell},\pm} = \frac{1}{\sqrt{2}} \left(
                                  \begin{array}{c}
                                    \phi_{\ell} \\
                                    \mp i \phi_{{\ell}-1} \\
                                  \end{array}
                                \right) , \label{mcc:evalueskp} \\
K_{+}, \, {\ell} = 0 : \quad \,\,\,\,\, E_{0} &=& 0 \, , \qquad \qquad \qquad \,\,
\psi_{0} = \left(
                                  \begin{array}{c}
                                    \phi_0 \\
                                    0 \\
                                  \end{array}
                                \right) , \label{mcc:evalueskp0}
\end{eqnarray}
where $\pm$ refer to the conduction and valence bands, respectively.
Equation~(\ref{mcc:evalueskp}) describes an electron
(plus sign) and a hole (minus sign) series of energy levels,
with prefactor $\Gamma = \sqrt{2} \hbar v/\lambda_B$ (\ref{mcc:lb}),
proportional to the square root of the magnetic field.
In addition,
there is a special level (\ref{mcc:evalueskp0}) fixed at zero energy
that arises from the presence of the lowering operator in the
Hamiltonian, $\pi \phi_{0}=0$. The corresponding eigenfunction $\psi_0$
has non-zero amplitude on the $A$ sublattice,
but its amplitude is zero on the $B$ sublattice.
The form (\ref{mcc:diracmf}) of the Hamiltonian $H_{1,K_{-}}$
at the second valley, $K_{-}$, shows that its spectrum
is degenerate with that at $K_{+}$, with
the role of the $A$ and $B$ sublattices reversed:
\begin{eqnarray}
K_{-}, \, {\ell} \geq 1 : \quad E_{{\ell},\pm} &=&
\pm \frac{\sqrt{2} \hbar v}{\lambda_B} \sqrt{{\ell}}  \, , \qquad
\psi_{{\ell},\pm} = \frac{1}{\sqrt{2}} \left(
                                  \begin{array}{c}
                                    \pm i \phi_{{\ell}-1} \\
                                    \phi_{\ell} \\
                                  \end{array}
                                \right) , \label{mcc:evalueskm} \\
K_{-}, \, {\ell} = 0 : \quad \,\,\,\,\, E_{0} &=& 0 \, , \qquad \qquad \qquad \,\,
\psi_{0} = \left(
                                  \begin{array}{c}
                                    0 \\
                                    \phi_0 \\
                                  \end{array}
                                \right) . \label{mcc:evalueskm0}
\end{eqnarray}
Thus, the eigenfunction $\psi_0$ of the zero-energy level
has zero amplitude on the $B$ sublattice at valley $K_{+}$
and zero amplitude on the $A$ sublattice at $K_{-}$.
If we take into account electronic spin, which contributes a twofold
degeneracy of the energy levels, as well as valley degeneracy,
then the Landau level spectrum of monolayer graphene consists of
fourfold-degenerate Landau levels.

\subsection{The integer quantum Hall effect in monolayer graphene}
\label{ss:mcc:qhemono}

\begin{figure}[t]
\centering
\includegraphics*[width=1.0\textwidth]{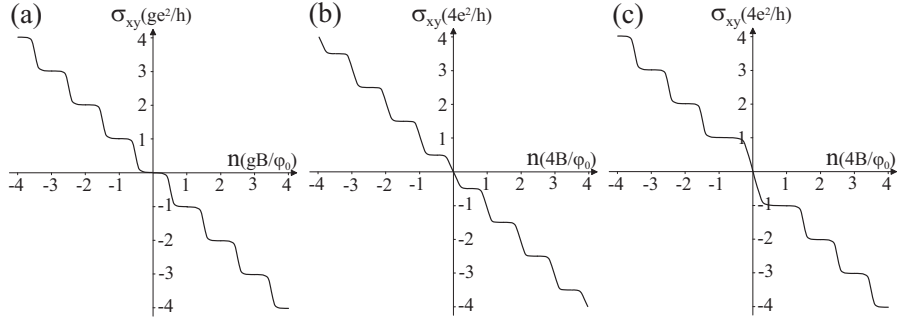}
\caption[]{Schematic representation of three types of integer
quantum Hall effect, showing the density dependence of the
Hall conductivity $\sigma_{xy}(n)$:
(a) conventional two-dimensional semiconductor systems with
additional system degeneracy $g$; (b) monolayer graphene; (c)
bilayer graphene.
Here, $B$ is the magnitude of the magnetic field
and $\varphi_0 = h/e$ is the flux quantum.}
\label{mccfig:16}
\end{figure}

In this section, we describe how the Landau level spectrum of graphene
is reflected in the dependence of the Hall conductivity $\sigma_{xy}(n)$
on carrier density $n$. In conventional two-dimensional semiconductor systems,
in the absence of any Berry's phase effects, the Landau level spectrum is
given by $E_{\ell} = \hbar \omega_c ({\ell} + 1/2)$, $\ell \geq 0$,
where $\omega_c = eB/m$ is the cyclotron frequency \cite{p+r87,macdonald89}.
Here, the lowest state lies at finite energy $E_{0} = \hbar \omega_c/2$.
If the system has an additional degeneracy $g$ (for example, $g=2$ for spin),
then plateaus \cite{vk80,p+r87,macdonald89}
occur at quantized $\sigma_{xy}$ values of $N (ge^2/h)$ where $N$ is an integer
and $e^2/h$ is the quantum value of conductance, i.e. each step between adjacent
plateaus has height $ge^2/h$, Fig.~\ref{mccfig:16}(a).
Each $\sigma_{xy}$ step coincides with the crossing of a Landau level on the
density axis.
Since the maximum carrier density per Landau level is $g B / \varphi_0$,
where $\varphi_0 = h/e$ is the flux quantum, the distance
between the $\sigma_{xy}$ steps on the density axis is $g B / \varphi_0$.

As described above, monolayer graphene has fourfold (spin and valley) degenerate
Landau levels $E_{{\ell},\pm} = \pm \sqrt{2{\ell}} \hbar v/\lambda_B$
for ${\ell} \geq 1$ and $E_{0} = 0$. The Hall conductivity $\sigma_{xy} (n)$,
Fig.~\ref{mccfig:16}(b),
displays a series of quantized plateaus separated by steps of size $4e^2/h$,
as in the conventional case, but the plateaus occur at half-integer values of
$4e^2/h$ rather than integer ones:
\begin{eqnarray}
\sigma_{xy} = - \frac{1}{2} \left(2N + 1 \right) \left( \frac{4e^2}{h} \right) \, ,
\end{eqnarray}
where $N$ is an integer, as observed experimentally \cite{novo05,zhang05}
and described theoretically \cite{haldane88,zheng02,gusynin05,peres06,herbut07}.
This unusual sequencing of $\sigma_{xy}$ plateaus is explained
by the presence of the fourfold-degenerate Landau level $E_0$ fixed at
zero energy. Since it lies at the boundary between the electron and hole gases,
it creates a step in $\sigma_{xy}$ of $4e^2/h$ at zero density.
Each Landau level in monolayer graphene is fourfold degenerate,
including the zero energy one, so the distance
between each $\sigma_{xy}$ step on the density axis is $4 B / \varphi_0$,
i.e. the steps occur at densities equal to integer values of $4 B / \varphi_0$.

\subsection{The Landau level spectrum of bilayer graphene}
\label{ss:mcc:llbi}

In the presence of a perpendicular magnetic field,
the Hamiltonian (\ref{mcc:bi1}) describing massive chiral electrons
in bilayer graphene may be written as
\begin{eqnarray}
\!\!\! \!\!\! H_{2,K_{+}} = - \frac{1}{2m} \left(
      \begin{array}{cc}
        0 & \left( \pi^{\dagger} \right)^2 \\
        \pi^2 & 0 \\
      \end{array}
    \right) ,
\qquad
H_{2,K_{-}} = - \frac{1}{2m} \left(
      \begin{array}{cc}
        0 & \pi^2 \\
        \left( \pi^{\dagger} \right)^2 & 0 \\
      \end{array}
    \right) ,
\label{mcc:bilayermf}
\end{eqnarray}
in the vicinity of corners of the Brillouin zone
$K_{+}$ and $K_{-}$, respectively.
Using the action of operators $\pi$ and $\pi^{\dagger}$ on the harmonic
oscillator states $\phi_{\ell}$, (\ref{mcc:lower}) and (\ref{mcc:raise}),
the Landau level spectrum of bilayer graphene may be calculated \cite{mcc06a}.
At the first valley, $K_{+}$, the Landau level energies and eigenstates
of $H_{2,K_{+}}$ are
\begin{eqnarray}
\!\!\! K_{+}, \, {\ell} \geq 2 : \quad E_{{\ell},\pm} &=&
\pm \frac{\hbar^2}{m\lambda_B^2} \sqrt{{\ell} ({\ell}-1)}  \, , \quad
\psi_{{\ell},\pm} = \frac{1}{\sqrt{2}} \left(
                                  \begin{array}{c}
                                    \phi_{\ell} \\
                                    \pm \phi_{{\ell}-2} \\
                                  \end{array}
                                \right) \! , \label{mcc:bikp} \\
\!\!\! K_{+}, \, {\ell} = 1 : \quad \,\,\,\,\, E_{1} &=& 0 \, , \quad \qquad \qquad \qquad \qquad
\psi_{1} = \left(
                                  \begin{array}{c}
                                    \phi_1 \\
                                    0 \\
                                  \end{array}
                                \right) , \label{mcc:bikp1} \\
\!\!\! K_{+}, \, {\ell} = 0 : \quad \,\,\,\,\, E_{0} &=& 0 \, , \quad \qquad \qquad \qquad \qquad
\psi_{0} = \left(
                                  \begin{array}{c}
                                    \phi_0 \\
                                    0 \\
                                  \end{array}
                                \right) , \label{mcc:bikp0}
\end{eqnarray}
where $\pm$ refer to the conduction and valence bands, respectively.
Equation~(\ref{mcc:bikp}) describes an electron
(plus sign) and a hole (minus sign) series of energy levels.
The prefactor $\hbar^2/(m\lambda_B^2)$ is proportional
to the magnetic field, and it may equivalently be written as
$\Gamma^2 / \gamma_1$ or as $\hbar \omega_c$ where $\omega_c = eB/m$.
For high levels, ${\ell} \gg 1$, the spectrum consists of approximately
equidistant levels with spacing $\hbar \omega_c$.
Note, however, that we are considering the low-energy Hamiltonian,
so that the above spectrum is only valid for sufficiently
small level index and magnetic field $\sqrt{{\ell}} \Gamma \ll \gamma_1$.
As well as the field-dependent levels, there are two special levels,
(\ref{mcc:bikp1}) and (\ref{mcc:bikp0}), fixed at zero energy.
There are two zero-energy levels because of the presence of the square of
the lowering operator in the Hamiltonian. It may act on the oscillator ground
state to give zero energy, $\pi^2 \phi_{0}=0$, (\ref{mcc:bikp0}),
but also on the first excited state to give zero energy, $\pi^2 \phi_{1}=0$,
(\ref{mcc:bikp1}).
The corresponding eigenfunctions $\psi_0$ and $\psi_1$
have non-zero amplitude on the $A1$ sublattice, that lies on the bottom
layer, but their amplitude is zero on the $B2$ sublattice.

The form (\ref{mcc:bilayermf}) of the Hamiltonian $H_{2,K_{-}}$
at the second valley, $K_{-}$, shows that its spectrum
is degenerate with that at $K_{+}$ with
the role of the $A1$ and $B2$ sublattices reversed.
It may be expressed as $H_{2,K_{-}} = \sigma_x H_{2,K_{+}} \sigma_x$
so that $\psi_{{\ell},\pm} (K_{-}) = \sigma_x \psi_{{\ell},\pm} (K_{+})$,
$\psi_{1} (K_{-}) = \sigma_x \psi_{1} (K_{+})$,
and $\psi_{0} (K_{-}) = \sigma_x \psi_{0} (K_{+})$.
Thus, the eigenfunctions $\psi_0$ and $\psi_1$ of the zero-energy levels
have zero amplitude on the $B2$ sublattice at valley $K_{+}$
and zero amplitude on the $A1$ sublattice at $K_{-}$.
If we take into account electronic spin, which contributes a twofold
degeneracy of the energy levels, as well as valley degeneracy,
then the Landau level spectrum of bilayer graphene consists of
fourfold degenerate Landau levels, except for the zero-energy levels
which are eightfold degenerate. This doubling of the degeneracy of the
zero-energy levels is reflected in the density dependence of
the Hall conductivity.

\subsection{The integer quantum Hall effect in bilayer graphene}
\label{ss:mcc:qhebi}

The Hall conductivity $\sigma_{xy} (n)$ of bilayer graphene,
Fig.~\ref{mccfig:16}(c),
displays a series of quantized plateaus occurring at integer values of
$4e^2/h$ that is practically the same as in the conventional case,
Fig.~\ref{mccfig:16}(a), with degeneracy per level $g=4$ accounting for spin and valleys.
However, there is a step of size $8e^2/h$ in $\sigma_{xy}$ across
zero density in bilayer graphene \cite{novo06,mcc06a}.
This unusual behavior is explained by the eightfold degeneracy of the
zero-energy Landau levels. Their presence creates a step in
$\sigma_{xy}$ at zero density, as in monolayer graphene,
but owing to the doubled degeneracy as compared to other levels,
it requires twice as many carriers to fill them. Thus,
the transition between the corresponding plateaus is twice as
wide in density, $8B/\varphi_0$ as compared to $4B/\varphi_0$, and the
step in $\sigma_{xy}$ between the plateaus must be twice as high,
$8e^2/h$ instead of $4e^2/h$.
This demonstrates that, although Berry's phase $2\pi$ is not reflected
in the sequencing of quantum Hall plateaus at high density,
it has a consequence in the quantum limit of zero density,
as observed experimentally \cite{novo06}.

Here, we showed that the chiral Hamiltonians of monolayer and bilayer graphene
corresponding to Berry's phase $\pi$ and $2\pi$, respectively,
have associated four- and eight-fold degenerate zero-energy Landau levels,
producing steps of four and eight times the conductance quantum $e^2/h$
in the Hall conductivity across zero density \cite{novo05,zhang05,novo06}.
In our discussion, we neglected interaction effects
and we assumed that any valley and spin splitting,
or splitting of the ${\ell}=1$ and ${\ell}=0$ levels in bilayer graphene,
are negligible as compared to temperature and level broadening.

\section{Trigonal warping in graphene}
\label{s:mcc:twg}

So far, we have described the tight-binding model of graphene
and showed that the low-energy Hamiltonians of monolayer and bilayer
graphene support chiral electrons with unusual properties.
There are, however, additional contributions
to the Hamiltonians that perturb this simple picture. In this section,
we focus on one of them, known in the graphite literature as trigonal warping
\cite{dre74,nak76,ino62,gup72,dressel02}.

\subsection{Trigonal warping in monolayer graphene}
\label{ss:mcc:twmono}

\begin{figure}[t]
\centering
\includegraphics*[width=0.6\textwidth]{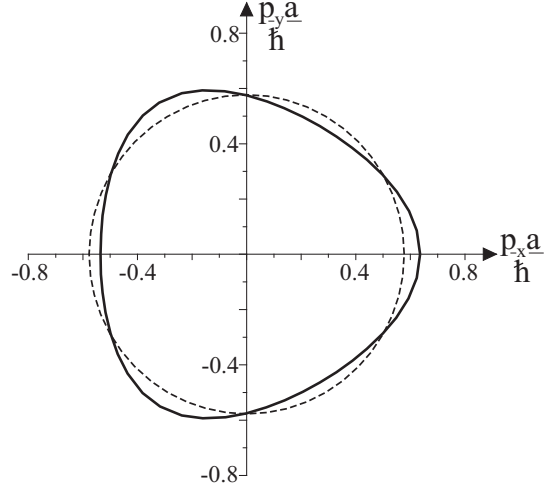}
\caption[]{Trigonal warping in monolayer graphene. The solid line
shows the isoenergetic line $E = 0.5 \gamma_0$ in the vicinity of
the valley $K_{+}$ using Eq.~(\ref{mcc:mtw}), the dashed line
shows the circular isoenergetic line obtained by
neglecting trigonal warping $\mu = 0$.}
\label{mccfig:14}
\end{figure}

The band structure of monolayer graphene, shown in Fig.~\ref{mccfig:17},
is approximately linear in the vicinity of zero energy, but it
shows deviations away from linear behavior at higher energy.
In deriving the Dirac-like Hamiltonian of monolayer graphene (\ref{mcc:dirac1}),
we kept only linear terms in the momentum $\mathbf{p} = \hbar \mathbf{k} - \hbar \mathbf{K}_{\xi}$
measured with respect to the $K_{\xi}$ point.
If we retain quadratic terms in $\mathbf{p}$, then the function $f ( \mathbf{k} )$,
(\ref{mcc:flinear}),
describing coupling between the $A$ and $B$ sublattices becomes
\begin{eqnarray}
f \left( \mathbf{k}\right) &\approx&
- \frac{\sqrt{3}a}{2\hbar} ( \xi p_x - i p_y)
 + \frac{a^2}{8\hbar^2} ( \xi p_x + i p_y)^2  , \label{mcc:fkexp}
\end{eqnarray}
where $p a / \hbar \ll 1$.
Using this approximate expression, the Dirac-like Hamiltonian (\ref{mcc:dirac1})
in the vicinity of point $K_{\xi}$ is modified \cite{ando98} as
\begin{eqnarray}
\!\!\! H_{1,\xi} = v \left(
      \begin{array}{cc}
        0 &   \xi p_x - i p_y  \\
         \xi p_x + i p_y  & 0 \\
      \end{array}
    \right)
- \mu
\left(
  \begin{array}{cc}
    0 & \left( \xi p_x + i p_y \right)^2 \\
    \left( \xi p_x - i p_y \right)^2 & 0 \\
  \end{array}
\right) \! ,
 \label{mcc:dirac3}
\end{eqnarray}
where parameter $\mu = \gamma_0 a^2 / (8\hbar^2)$.
The corresponding energy eigenvalues are
\begin{eqnarray}
E_{\pm} = \pm \sqrt{v^2p^2 - 2 \xi \mu v p^3 \cos 3 \varphi + \mu^2 p^4} \, , \label{mcc:mtw}
\end{eqnarray}
For small momentum near the $K$ point, $p a / \hbar \ll 1$,
the terms containing parameter $\mu$ are a small perturbation
because $\mu p^2 / (v p) = p a / (4 \sqrt{3} \hbar)$.
They contribute to a weak triangular deformation of the Fermi circle
that becomes stronger as the momentum $\mathbf{p}$ becomes larger.
Figure~\ref{mccfig:14} shows the trigonal warping of the Fermi
circle near point $K_{+}$, obtained by plotting (\ref{mcc:mtw})
for constant energy $E = 0.5 \gamma_0$.
The presence of the valley index $\xi = \pm 1$ in the angular
term of (\ref{mcc:mtw}) means that the orientation of the trigonal warping
at the second valley $K_{-}$ is reversed.

\subsection{Trigonal warping and Lifshitz transition in bilayer graphene}
\label{ss:mcc:twbi}

In deriving the low-energy Hamiltonian of bilayer graphene (\ref{mcc:bi1}),
the linear approximation of $f ( \mathbf{k} )$ (\ref{mcc:flinear}) in the
vicinity of the $K$ point was used. Taking into account quadratic
terms in $f ( \mathbf{k} )$ would produce higher-order in momentum
contributions to (\ref{mcc:bi1}), that would tend to be relevant
at large momentum $\mathbf{p}$.
There is, however, an additional interlayer coupling in bilayer graphene
that contributes to trigonal warping and tends to be relevant at
small momentum $\mathbf{p}$, {i.e.} at low energy and very close to
the $K$ point.

\begin{figure}[t]
\centering
\includegraphics*[width=0.5\textwidth]{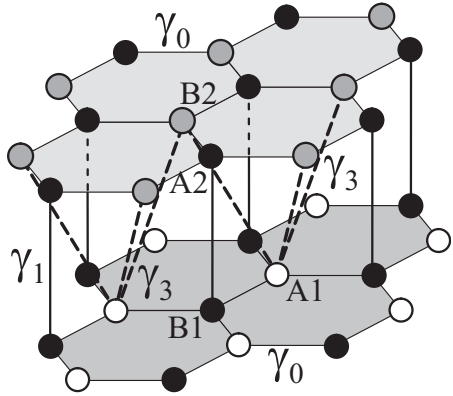}
\caption[]{Schematic representation of the crystal structure of
$AB$-stacked bilayer graphene illustrating skew interlayer coupling $\gamma_3$ (dashed lines) between $p_z$ orbitals on sites $A1$ (white) and $B2$ (grey)}
\label{mccfig:15}
\end{figure}

The additional coupling is a skew interlayer coupling
between $p_z$ orbitals on atomic sites $A1$ and $B2$,
Fig.~\ref{mccfig:15}, denoted $\gamma_3$. For each $A1$ site,
there are three $B2$ sites nearby.
A calculation of the matrix element between $A1$ and $B2$ sites in
the tight-binding model
proceeds in a similar way as that between adjacent $A$ and $B$ sites in
monolayer graphene, as described in Sect.~\ref{ss:mcc:offdiag}.
Then, the effective Hamiltonian in a basis with components $A1$, $B1$, $A2$, $B2$,
for the four low-energy bands of bilayer graphene (\ref{mcc:Hbifull2}) is \cite{mcc06a}:
\begin{eqnarray}
H = \left(
      \begin{array}{cccc}
        0 & - \gamma_0 f \left( \mathbf{k} \right) & 0 & - \gamma_3 f^{\ast} \left( \mathbf{k} \right) \\
        - \gamma_0 f^{\ast} \left( \mathbf{k} \right) & 0 & \gamma_1 & 0 \\
        0 & \gamma_1 & 0 & - \gamma_0 f \left( \mathbf{k} \right) \\
        - \gamma_3 f \left( \mathbf{k} \right) & 0 & - \gamma_0 f^{\ast} \left( \mathbf{k} \right) & 0 \\
      \end{array}
    \right) \,  , \label{mcc:Hbifull3}
\end{eqnarray}
where
\begin{eqnarray}
\gamma_3 =
- \langle \phi_{A1} \left( \mathbf{r} - \mathbf{R}_{A1} \right) | {\cal H} | \phi_{B2} \left( \mathbf{r} - \mathbf{R}_{B2} \right) \rangle \, .
\end{eqnarray}
The $\gamma_3$ term is relevant at low energy because it is
a direct coupling between the $A1$ and $B2$ orbitals that form
the two low-energy bands.
Thus, using the linear-in-momentum approximation (\ref{mcc:flinear}), terms such as
$\gamma_3 f \left( \mathbf{k} \right) \approx - v_3 \left( \xi p_x - i p_y \right)$
appear in the two-component Hamiltonian written in
basis ${c}_{A1}$, ${c}_{B2}$. Equation~(\ref{mcc:bi1}) is modified
as \cite{mcc06a}
\begin{eqnarray}
\!\!\!\!\!\! H_{2,\xi} = v_3 \! \left(
      \begin{array}{cc}
        0 &   \xi p_x + i p_y  \\
         \xi p_x - i p_y  & 0 \\
      \end{array}
    \right) \! - \!\frac{1}{2m}\! \left(
      \begin{array}{cc}
        0 & \left( \xi p_x - i p_y \right)^2 \\
        \left( \xi p_x + i p_y \right)^2 & 0 \\
      \end{array}
    \right) \!\! , \label{mcc:bi3}
\end{eqnarray}
where $v_3 = \sqrt{3} a \gamma_3/ (2 \hbar )$ and $m = \gamma_1 / (2v^2)$.
Taking into account trigonal warping, the low-energy Hamiltonian
of bilayer graphene (\ref{mcc:bi3}) resembles that of monolayer
graphene (\ref{mcc:dirac3}). The principle difference lies in
the magnitude of the parameters. Since $\gamma_3 = 0.315\,$eV \cite{dressel02}
is an order of magnitude less than $\gamma_0 = 3.033\,$eV \cite{saito},
then $v_3 \ll v$. Thus, the linear term dominates in monolayers and
the quadratic term dominates in bilayers over a broad range of energy.
The energy eigenvalues of $H_{2,\xi}$, (\ref{mcc:bi3}), are
\begin{eqnarray}
E_{\pm} = \pm \sqrt{v_3^2 p^2 - \xi \frac{v_3 p^3}{m} \cos 3 \varphi
+ \left(\frac{p^2}{2m}\right)^2 } \, , \label{mcc:bitw}
\end{eqnarray}
for energies $|E_{\pm}| \ll \gamma_1$. Over a range of energy,
the term independent of $v_3$ dominates, and
the $v_3$ dependent terms produce trigonal warping of
the isoenergetic line in the vicinity of each $K$ point.
The effect of trigonal warping increases as the energy is lowered,
until, at very low energies $E_L \approx
{\textstyle\frac{1}{4}}\gamma_{1} ( v_{3} / v )^{2} \approx 1\,$meV,
it leads to a Lifshitz transition \cite{lif60}: the isoenergetic line breaks
into four parts \cite{dre74,nak76,ino62,gup72,dressel02,mcc06a,part06,mcc07}.
There is one `central' part, centered on the $K$ point ($p=0$),
that is approximately circular with area
$\mathcal{A}_{\mathrm{c}} \approx \pi E^{2}/(\hbar v_{3})^{2}$.
In addition, there are three `leg' parts that are elliptical with area
$\mathcal{A}_{\mathrm{\ell}} \approx
{\textstyle\frac{1}{3}}\mathcal{A}_{\mathrm{c}}$.
Each ellipse has its major axis separated
by angle $2\pi/3$ from the major axes of the other leg parts,
as measured from the $K$ point, with the
ellipse centered on $|p|=\gamma _{1}v_{3}/v^{2}$.

Here, we have described the low-energy band structure of monolayer and
bilayer graphene within a simple tight-binding model, including
a Lifshitz transition in bilayer graphene at very low energy $E_L \approx 1\,$meV.
It is quite possible that electron-electron interactions
have a dramatic effect on the band structure of bilayer graphene,
producing qualitatively different features at low energy
\cite{nil06,min08,zhangmin09,nand09,vafek10,lem10}.

\section{Tuneable band gap in bilayer graphene}
\label{s:mcc:tune}

\subsection{Asymmetry gap in the band structure of bilayer graphene}
\label{ss:mcc:gap}

In graphene monolayers and bilayers, a combination of space and time
inversion symmetry \cite{manes07} guarantees the existence of a gapless band structure
exactly at the $K$ point, {i.e.} the $A$ and $B$ sublattices
($A1$ and $B2$ in bilayers) are identical, leading to degeneracy of
the states they support at the $K$ point. Breaking inversion symmetry
by, say, fixing the two sublattice sites to be at different energies,
would lead to a gap between the conduction and valence bands at the
$K$ point. In monolayer graphene, breaking the $A$/$B$ sublattice symmetry
in a controllable way is very difficult: it would require a periodic
potential because $A$ and $B$ are adjacent
sites on the same layer. In bilayer graphene, however,
the $A1$ and $B2$ sublattices lie on different layers and, thus,
breaking the symmetry and opening a band gap may be achieved by doping or
gating.
Band-gap opening in bilayer graphene has recently been studied
both theoretically \cite{mcc06a,latil06,guinea06,mcc06b,min07,aoki07,gava,castro,zhang08,fogler10}
and in a range of different experiments
\cite{ohta06,oostinga,castro,zhang08,hen08,li08,zhang09,mak09,kuz09a,kuz09b,kuz09c,zhao10,kim10}.

\begin{figure}[t]
\centering
\includegraphics*[width=.7\textwidth]{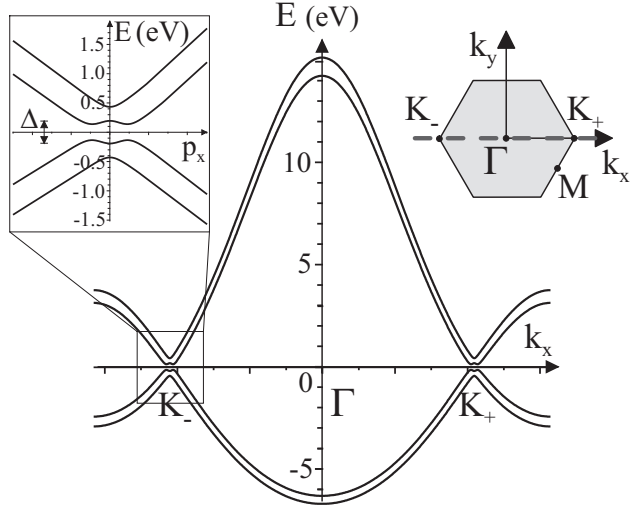}
\caption[]{The low-energy band structure of bilayer
graphene in the presence of interlayer asymmetry $\Delta$.
Parameter values are $\gamma_0 = 3.033\,$eV, $s_0 = 0.129$,
$\Delta = \gamma_1 = 0.39\,$eV.
The plot shows the bands calculated along the $k_x$ axis intersecting
points $K_{-}$, $\Gamma$, and $K_{+}$ in the Brillouin zone,
shown as the dotted line in the right inset. The left inset shows the
band structure in the vicinity of the point $K_{-}$.}
\label{mccfig:10}
\end{figure}

If we introduce an asymmetry parameter $\Delta = \epsilon_{2} - \epsilon_{1}$
describing the difference between on-site energies in the two layers,
$\epsilon_{A2} = \epsilon_{B2} = \epsilon_{2} = {\textstyle\frac{1}{2}} \Delta$,
$\epsilon_{A1} = \epsilon_{B1} = \epsilon_{1} = -{\textstyle\frac{1}{2}} \Delta$,
then the transfer integral matrix of bilayer graphene (\ref{mcc:Hbifull}),
in a basis with components $A1$, $B1$, $A2$, $B2$, becomes \cite{mcc06a,guinea06,mcc06b}
\begin{eqnarray}
H &=& \left(
      \begin{array}{cccc}
        -{\textstyle\frac{1}{2}} \Delta & - \gamma_0 f \left( \mathbf{k} \right) & 0 & 0 \\
        - \gamma_0 f^{\ast} \left( \mathbf{k} \right) & -{\textstyle\frac{1}{2}} \Delta & \gamma_1 & 0 \\
        0 & \gamma_1 & {\textstyle\frac{1}{2}} \Delta & - \gamma_0 f \left( \mathbf{k} \right) \\
        0 & 0 & - \gamma_0 f^{\ast} \left( \mathbf{k} \right) & {\textstyle\frac{1}{2}} \Delta \\
      \end{array}
    \right) \,  , \label{mcc:HbifullD}
\end{eqnarray}
The band structure may be determined by solving the secular equation
$\det \left( H -  E_j S \right) = 0$ using overlap matrix $S$, (\ref{mcc:Sbifull}).
It is plotted in Fig.~\ref{mccfig:10} for parameter values
$\gamma_0 = 3.033\,$eV, $s_0 = 0.129$ and $\Delta = \gamma_1 = 0.39\,$eV.
A band gap appears between the conduction and valence bands near the $K$ points (left
inset in Fig.~\ref{mccfig:10}).

To develop an analytic description of the bands at low energy,
we neglect non-orthogonality of the orbitals on adjacent sites,
so that the overlap matrix $S$, (\ref{mcc:Sbifull}), becomes a unit matrix.
Then, the bands at low energy are described by Hamiltonian, (\ref{mcc:HbifullD}),
with eigenvalues \cite{mcc06a} given by
\begin{eqnarray}
E_{\pm}^{(\alpha)} = \pm \left[ \frac{\Delta^2}{4} + v^2p^2
+ \frac{\gamma_1^2}{2} + \alpha \frac{\gamma_1^2}{2}
\sqrt{1 + \frac{4v^2p^2}{\gamma_1^2}
+ \frac{4 \Delta^2v^2p^2}{\gamma_1^4}} \right]^{1/2} \!\!\! \!\!\! , \label{mcc:EDel}
\end{eqnarray}
where $\alpha = 1$ for the split bands and $\alpha = -1$ for the low-energy bands.
Here, we used the linear approximation
$f \left( \mathbf{k} \right) \approx - v \left( \xi p_x - i p_y \right)/\gamma_0$,
(\ref{mcc:flinear}), so that $\gamma_0 |f\left( \mathbf{k} \right)| \approx vp$.
Eigenvalues $E_{\pm}^{(-1)}$ describe the low-energy bands split by
a gap. They have a distinctive `Mexican hat' shape, shown in the left
inset in Fig.~\ref{mccfig:10}. The separation between the bands
exactly at the $K$ point, $E_{+}^{(-1)} (p=0) - E_{-}^{(-1)} (p=0)$,
is equal to $|\Delta|$, but the true value of the band gap $\Delta_g$ occurs at non-zero value of
the momentum $p_g$ away from the $K$ point,
\begin{eqnarray}
p_g &=& \frac{|\Delta|}{2v} \sqrt{\frac{\Delta^2 + 2 \gamma_1^2}{\Delta^2 + \gamma_1^2}} \, , \\
\Delta_g &=& E_{+}^{(-1)} (p_g) - E_{-}^{(-1)} (p_g) =
\frac{|\Delta|\gamma_1}{\sqrt{\Delta^2 + \gamma_1^2}} \, . \label{mcc:dg}
\end{eqnarray}
For moderate values of the asymmetry parameter, $|\Delta| \ll \gamma_1$,
then the band gap $\Delta_g \approx |\Delta|$, but for extremely
large values, $|\Delta| \gg \gamma_1$, the gap saturates $\Delta_g \approx \gamma_1$,
where $\gamma_1$ is of the order of three to four hundred meV.
The value of the asymmetry parameter $\Delta$ and bandgap $\Delta_g$ may
be tuned using an external gate potential, but the ability of an external
gate to induce a potential asymmetry between the layers of the bilayer
depends on screening by the electrons in bilayer graphene, as discussed in
the following.

\subsection{Self-consistent model of screening in bilayer graphene}
\label{ss:mcc:screen}

\subsubsection{Introduction}

The influence of screening on band-gap opening in bilayer graphene
has been modeled using the tight-binding model and Hartree theory
\cite{mcc06b,min07,castro,zhang08,fogler10}, and this simple analytic model is in good
qualitative agreement with density functional theory \cite{min07,gava}
and experiments \cite{castro,li08,zhang08,mak09,kuz09b}.
Recently, the tight-binding model and Hartree theory approach
has been applied to graphene trilayers and multilayers \cite{kosh09a,avet,avet2,kosh10a}.
Here, we review the tight-binding model and Hartree theory approach
which provides analytical formulae that serve to illustrate the pertinent physics.
We will use the SI system of units throughout, and adopt the convention
that the charge of the electron is $-e$ where the quantum of charge $e > 0$.

Using elementary electrostatics, it is possible to relate the asymmetry parameter
$\Delta = \epsilon_{2} - \epsilon_{1}$
to the distribution of electronic density over the bilayer system
in the presence of external gates,
but the density itself depends explicitly on $\Delta$
because of the effect $\Delta$ has on the band structure, (\ref{mcc:EDel}).
Therefore, the problem requires a self-consistent calculation of density and $\Delta$,
leading to a determination of
the gate-dependence of the gap $\Delta_g$.

\begin{figure}[t]
\centering
\includegraphics*[width=.8\textwidth]{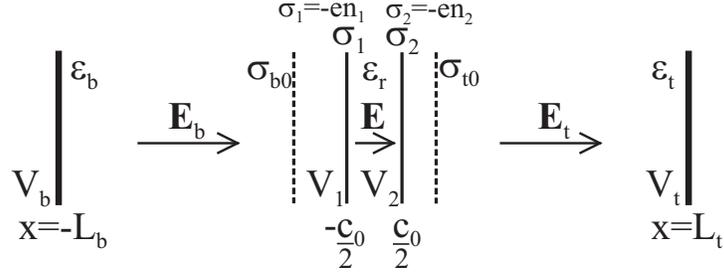}
\caption[]{Schematic of bilayer graphene in the presence of
back and top gates.
Bilayer graphene consists of two parallel conducting plates
with respective electron densities
$n_1$, $n_2$ located at $x=-c_0/2$ and $+c_0/2$, respectively,
where $c_0$ is the interlayer spacing, and $\varepsilon_r$ is
the permittivity of the bilayer interlayer space.
The back (top) gate at $x=-L_b$ ($x= +L_t$), held at potential $V_b$ ($V_t$), is
separated from the bilayer by a dielectric medium with relative
permittivity $\varepsilon_b$ ($\varepsilon_t$).
Dashed lines indicate additional background charge near
the bilayer with charge densities $\sigma_{b0}$ and $\sigma_{t0}$
on the back-gate and top-gate side, respectively.}
\label{mccfig:11}
\end{figure}

The model assumes that bilayer graphene consists of two parallel conducting plates
located at $x=-c_0/2$ and $+c_0/2$, where $c_0$ is the interlayer spacing,
as illustrated in Fig.~\ref{mccfig:11}. The two layers support
electron densities $n_1$, $n_2$, respectively, corresponding to charge
densities $\sigma_1 = -e n_1$, $\sigma_2 = -e n_2$, and
the permittivity of the bilayer interlayer space is $\varepsilon_r$ (neglecting
the screening effect of $\pi$-band electrons that we
explicitly take into account here).
We consider the combined effect of a back and top gate, with
the back (top) gate at $x=-L_b$ ($x= +L_t$), held at potential $V_b$ ($V_t$),
separated from the bilayer by a dielectric medium with relative
permittivity $\varepsilon_b$ ($\varepsilon_t$).
In addition, we include the influence of additional background charge near
the bilayer with density $n_{b0}$ on the back-gate side and $n_{t0}$
on the top-gate side,
yielding charge densities $\sigma_{b0} = e n_{b0}$ and $\sigma_{t0} = e n_{t0}$
where $n_{b0}$ and $n_{t0}$ are positive for positive charge.

\subsubsection{Electrostatics}

Applying Gauss's Law firstly to a Gaussian surface enclosing cross-sectional
area $A$ of both layers of the bilayer and, secondly, to a Gaussian surface
enclosing one layer only yields
\begin{eqnarray}
- \varepsilon_0 \varepsilon_b E_b A + \varepsilon_0 \varepsilon_t E_t A
&=& - e \left( n_1 + n_2 - n_{b0} - n_{t0} \right) A \, , \label{mcc:EA}\\
- \varepsilon_0 \varepsilon_r E A + \varepsilon_0 \varepsilon_t E_t A
&=& - e \left( n_2 - n_{t0} \right) A \, . \label{mcc:EB}
\end{eqnarray}
The electric fields may be related to potential differences,
\begin{eqnarray}
E_b &\approx& V_b/L_b \, , \qquad E_t \approx - V_t/L_t \, , \\
E &\approx& \left(V_1 - V_2\right)/c_0 \equiv \Delta / (ec_0) \, .
\end{eqnarray}
and, when substituted into (\ref{mcc:EA}) and (\ref{mcc:EB}), they give
\begin{eqnarray}
n = n_1 + n_2
&=& \frac{\varepsilon_0 \varepsilon_b V_b}{eL_b}
+ \frac{\varepsilon_0 \varepsilon_t V_t}{eL_t}
+ n_{b0} + n_{t0} \, , \label{mcc:VA} \\
\Delta
&=& - \frac{\varepsilon_t}{\varepsilon_r}\frac{c_0}{L_t}e V_t + \frac{e^2 c_0}{\varepsilon_0 \varepsilon_r} \left( n_2 - n_{t0} \right) \, . \label{mcc:VB}
\end{eqnarray}
The first equation, (\ref{mcc:VA}), relates the total density of
$\pi$-band electrons $n = n_1 + n_2$ on the bilayer
to the gate potentials, generalizing the case of monolayer graphene \cite{novo04}.
The second equation, (\ref{mcc:VB}), gives the value of the asymmetry
parameter. Using (\ref{mcc:VA}), it may be written in a slightly different way:
\begin{eqnarray}
\Delta &=& \Delta_{\mathrm{ext}}
+ \Lambda \gamma_1 \frac{\left( n_2 - n_1 \right)}{n_{\perp}} \, , \label{mcc:VC} \\
\Delta_{\mathrm{ext}}
&=& \frac{1}{2}\frac{\varepsilon_b}{\varepsilon_r}\frac{c_0}{L_b}e V_b
- \frac{1}{2}\frac{\varepsilon_t}{\varepsilon_r}\frac{c_0}{L_t}e V_t
+ \Lambda \gamma_1 \frac{\left( n_{b0} - n_{t0} \right)}{n_{\perp}}
\, , \label{mcc:VD}
\end{eqnarray}
where parameters $n_{\perp}$ and $\Lambda$ are defined as
\begin{eqnarray}
n_{\perp} = \frac{\gamma_1^2}{\pi \hbar^2v^2} \, , \qquad
\Lambda = \frac{c_0 e^2 \gamma_1}{2\pi \hbar^2v^2 \varepsilon_0 \varepsilon_r}
\equiv \frac{c_0 e^2 n_{\perp}}{2 \gamma_1 \varepsilon_0 \varepsilon_r} \, .
\end{eqnarray}
The first term in (\ref{mcc:VC}) is $\Delta_{\mathrm{ext}}$,
the value of $\Delta$ if screening were negligible,
as determined by a difference between the gate potentials, (\ref{mcc:VD}).
Equations~(\ref{mcc:VA},\ref{mcc:VD}) show that the effect of the background
densities $n_{b0}$ and $n_{t0}$ may be absorbed in a shift of the gate
potentials $V_b$ and $V_t$, respectively.

The second term in (\ref{mcc:VC}) indicates the influence of screening by electrons on the bilayer where $n_{\perp}$ is the characteristic density scale
and $\Lambda$ is a dimensionless parameter indicating
the strength of interlayer screening. Using $\gamma_1 = 0.39$eV
and $v = 1.0 \times 10^6$ms$^{-1}$
gives $n_{\perp} = 1.1 \times 10^{13}$cm$^{-2}$.
For interlayer spacing $c_0 = 3.35${\AA} and dielectric constant $\varepsilon_r \approx 1$,
then $\Lambda \sim 1$, indicating that screening is an important effect.

\subsubsection{Layer densities}

Equation~(\ref{mcc:VC}) uses electrostatics to relate $\Delta$ to the
electronic densities $n_{1}$ and $n_{2}$ on the individual
layers.
The second ingredient of the self-consistent analysis are expressions
for $n_{1}$ and $n_{2}$ in terms of $\Delta$, taking into account the electronic band
structure of bilayer graphene.
The densities are determined by an integral with respect to momentum over the
circular Fermi surface
\begin{eqnarray}
n_{1(2)} = \frac{2}{\pi \hbar^2} \int p \, dp
\left[ | \psi_{A1(2)} ( p ) |^2 + | \psi_{B1(2)} ( p ) |^2 \right]
\end{eqnarray}
where a factor of four is included to take into
account spin and valley degeneracy.
Using the four-component Hamiltonian (\ref{mcc:HbifullD}),
with linear approximation
$f \left( \mathbf{k} \right) \approx - v \left( \xi p_x - i p_y \right)/\gamma_0$,
it is possible to determine the
wave function amplitudes on the four atomic sites \cite{mcc06b} to find
\begin{eqnarray}
\!\!\! n_{1(2)} &=&  \int d p  \, p \!
\left( \frac{ E \mp \Delta/2}{\pi \hbar^2 E} \right) \!\!\!
\left[ \frac{ \left( E^2 - \Delta^2
/4 \right)^2 \mp 2 v^2 p^2 E \Delta - v^4 p^4 } { \left(
E^2 - \Delta^2 /4 \right)^2 + v^2 p^2 \Delta^2 - v^4 p^4 }
\right] \!\! ,
\end{eqnarray}
where the minus (plus) sign is for the first (second) layer
and $E$ is the band energy.

For simplicity, we consider the Fermi
level to lie within the lower conduction band, but above
the Mexican hat region, $|\Delta|/2 < E_F \ll \gamma_1$.
We approximate the dispersion relation, (\ref{mcc:EDel}),
as $E_{+}^{(-1)} \approx \sqrt{\Delta^2/4 + v^4p^4/\gamma_1^2}$
which neglects features related to the Mexican hat.
Then, the contribution to the layer densities from the
partially-filled conduction band \cite{mcc06b,fogler10}
is given by
\begin{eqnarray}
n_{1(2)}^{\mathrm{cb}} \approx \frac{n}{2}
\mp \frac{n_{\perp} \Delta}{4\gamma_1} \ln
\left( \frac{2 |n| \gamma_1}{n_{\perp}|\Delta|}
+ \sqrt{1 + \left( \frac{2 n\gamma_1}{n_{\perp} \Delta} \right)^2} \right) \, ,
\end{eqnarray}
where the total density $n = p_F^2 / \pi \hbar^2$.
In addition, although the filled valence band doesn't contribute
to a change in the total density $n$, it contributes towards the finite
layer polarization in the presence of finite $\Delta$ which,
to leading order in $\Delta$, is given by
\begin{eqnarray}
n_{1(2)}^{\mathrm{vb}} \approx
\pm \frac{n_{\perp} \Delta}{4\gamma_1} \ln
\left( \frac{4\gamma_1}{|\Delta|} \right) \, .
\end{eqnarray}
Then, the total layer density,
$n_{1(2)} = n_{1(2)}^{\mathrm{cb}} + n_{1(2)}^{\mathrm{vb}}$, is given by
\begin{eqnarray}
n_{1(2)} \approx \frac{n}{2}
\mp \frac{n_{\perp} \Delta}{4\gamma_1} \ln
\left( \frac{|n|}{2n_{\perp}} +
\frac{1}{2}\sqrt{ \left( \frac{n}{n_{\perp}} \right)^2 + \left( \frac{\Delta}{2\gamma_1} \right)^2} \right) \, . \label{mcc:n12}
\end{eqnarray}

\subsubsection{Self-consistent screening}

The density-dependence of the asymmetry parameter $\Delta$
and band gap $\Delta_g$ are determined \cite{mcc06b,fogler10} by substituting
the expression for the layer density, (\ref{mcc:n12}),
into (\ref{mcc:VC}):
\begin{eqnarray}
\Delta \left( n \right) \approx \Delta_{\mathrm{ext}} \left[
1 - \frac{\Lambda}{2}
\ln
\left( \frac{|n|}{2n_{\perp}} +
\frac{1}{2}\sqrt{ \left( \frac{n}{n_{\perp}} \right)^2 + \left( \frac{\Delta}{2\gamma_1} \right)^2} \right)
\right]^{-1} \, , \label{mcc:delta}
\end{eqnarray}
with $\Delta_{\mathrm{ext}}$ given by (\ref{mcc:VD}).
The logarithmic term describes the influence of screening:
when this term is much smaller than unity, screening is negligible
and $\Delta \approx \Delta_{\mathrm{ext}}$, whereas when
the logarithmic term is much larger than unity,
screening is strong, $|\Delta| \ll |\Delta_{\mathrm{ext}}|$.
The magnitude of the logarithmic term is proportional to
the screening parameter $\Lambda$.
As discussed earlier, $\Lambda \sim 1$
in bilayer graphene, so it is necessary to take account
of the density dependence of the logarithmic term in (\ref{mcc:delta}).

\begin{figure}[t]
\centering
\includegraphics*[width=1.0\textwidth]{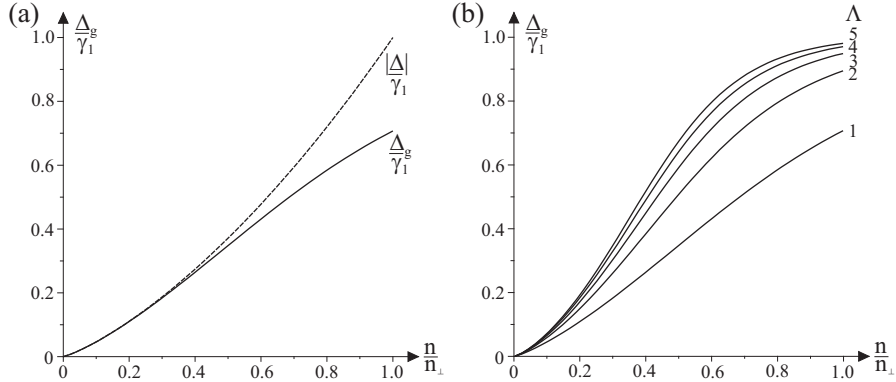}
\caption[]{Density-dependence of the band gap $\Delta_g$ in bilayer graphene,
in the presence of a single back gate:
(a) asymmetry parameter $\Delta$ and gap
$\Delta_g = |\Delta|\gamma_1/\sqrt{\Delta^2 + \gamma_1^2}$ for screening
parameter $\Lambda = 1$;
(b) band gap $\Delta_g$ for different values of the screening
parameter.
Plots were made using Eqs.~(\ref{mcc:dg}) and (\ref{mcc:delta2}).}
\label{mccfig:12}
\end{figure}

To understand the density dependence of $\Delta$, let us consider
bilayer graphene in the presence of a single back gate,
$V_t = n_{b0} = n_{t0} = 0$. This is a common situation for experiments
with exfoliated graphene on a silicon substrate
\cite{novo04,novo05,zhang05,novo06}.
Then, the relation between density and gate voltage, (\ref{mcc:VA}),
becomes the same as in monolayer graphene \cite{novo04},
$n = \varepsilon_0 \varepsilon_b V_b/(eL_b)$.
The expression for $\Delta_{\mathrm{ext}}$, (\ref{mcc:VD}),
reduces to $\Delta_{\mathrm{ext}} = \Lambda \gamma_1 n / n_{\perp}$,
and the expression for $\Delta \left( n \right)$, (\ref{mcc:delta}),
simplifies \cite{mcc06b} as
\begin{eqnarray}
\Delta \left( n \right) \approx \frac{\Lambda \gamma_1 n}{n_{\perp}}
\left[ 1 - \frac{\Lambda}{2}
\ln \left( \frac{|n|}{n_{\perp}} \right) \right]^{-1} \, , \label{mcc:delta2}
\end{eqnarray}
The value of the true band gap $\Delta_g (n)$ may be obtained using (\ref{mcc:dg}),
$\Delta_g = |\Delta|\gamma_1/\sqrt{\Delta^2 + \gamma_1^2}$.
Asymmetry parameter $\Delta (n)$ and band gap $\Delta_g (n)$ are plotted in Fig.~\ref{mccfig:12}
as a function of density $n$.
For large density, $|n| \sim n_{\perp}$, the logarithmic term in (\ref{mcc:delta2})
is negligible and the asymmetry parameter is approximately linear in density,
$\Delta (n) \approx \Lambda \gamma_1 n/n_{\perp}$.
At low density, $|n| \ll n_{\perp}$, the logarithmic term is large, indicating
that screening is strong, and the asymmetry parameter approaches
$\Delta (n) \approx 2 \gamma_1 (n/n_{\perp}) / \ln ( n_{\perp}/|n| )$.
The comparison of $\Delta (n)$ and $\Delta_g (n)$, Fig.~\ref{mccfig:12}(a),
shows that, at low density $|n| \ll n_{\perp}$, $\Delta_g (n) \approx |\Delta (n)|$
and, asymptotically, $\Delta_g (n) \approx 2 \gamma_1 (|n|/n_{\perp}) / \ln ( n_{\perp}/|n| )$.
This is independent of the screening parameter $\Lambda$
[see Fig.~\ref{mccfig:12}(b)].
The curves for different values
of the screening parameter $\Lambda$, Fig.~\ref{mccfig:12}(b), illustrate
that, even when $|\Delta|$ is very large, $|\Delta| \gg \gamma_1$, $\Delta_g$
saturates at the value of $\gamma_1$.

In deriving the above expression for $\Delta \left( n \right)$,
a number of approximations were made including
simplifying the band structure [by omitting features related to the
Mexican hat or to other possible terms in the Hamiltonian (\ref{mcc:HbifullD})],
neglecting screening due to other orbitals,
and neglecting the effects of disorder and electron-electron
exchange and correlation. Nevertheless, it seems to be in good qualitative
agreement with density functional theory calculations \cite{min07,gava}
and experiments (see, for example, \cite{castro,li08,zhang08,mak09,kuz09b}).

\section{Summary}
\label{s:mcc:summary}

In this Chapter, some of the electronic properties of monolayer and
bilayer graphene were described using the tight-binding model.
Effective Hamiltonians for low-energy electrons were derived,
corresponding to massless chiral fermions in monolayers and
massive chiral fermions in bilayers.
Chirality in graphene is manifest in many electronic properties,
including anisotropic scattering and an unusual sequence of
plateaus in the quantum Hall effect.
There are a number of additional contributions to the low-energy
Hamiltonians of graphene that influence chiral electrons and
we focused on one of them, trigonal warping, here.

Comparison with experiments suggest that the tight-binding model generally
works very well in graphene. The model contains parameters, corresponding
to the energies of atomic orbitals or to matrix elements describing hopping
between atomic sites, that cannot be determined by the model. They must be
estimated by an alternative theoretical method, such as
density-functional theory, or they can be treated as fitting parameters
to be determined by comparison with experiments.
The simple model described in this Chapter is versatile and it
serves as the starting point for a wide range of models encapsulating advanced
physical phenomena, including interaction effects, and
the tight-binding model may be used to describe the electronic structure
of multilayer graphene, too. Here, we described
a different example: the use of the tight-binding model with Hartree theory
to develop a simple model of screening by electrons in bilayer graphene in order
to calculate the density dependence of the band gap induced by an external
electric field.

\subsubsection{Acknowledgments}

The author thanks colleagues for fruitful collaboration in graphene
research, in particular V.I. Fal'ko, and EPSRC for financial support.



\printindex
\end{document}